\documentclass[a4paper,11pt]{article}
\usepackage{jheppub} 
\usepackage{lineno}

\title{\boldmath Sparse Partial-Tracing}

\author{Julio Candanedo}
\affiliation{Department of Physics, Arizona State University \\ Tempe, AZ 85287, USA}

\emailAdd{jcandane@asu.edu}

\abstract{Matrices and more generally multidimensional \textit{arrays}, form the backbone of computational studies.
In this paper we demonstrate increases in computational efficiency by performing partial-tracing/tensor-contractions on sparse-arrays.
It was shown that sparse-arrays are really 3 dense-arrays (dense-shape, index-array, and data-array).
Dense-array manipulations of these constituent arrays are used to determine the resulting partial-trace.
Because computational arrays are used in a verity of different studies, these methods are broadly applicable. }

\newcommand{\I}{\mathbb{I}}
\newcommand{\J}{\mathbb{J}}
\newcommand{\N}{\mathbb{N}}
\newcommand{\Z}{\mathbb{Z}}
\newcommand{\bipartnn}{\begin{tikzpicture}
    \SetVertexMath

    \foreach \y [count=\n,evaluate={\m=int(10-\n)}] in {4,...,-4}
    {\Vertex[x=6, y=\y, L=A_\n]{l2\m}
    }

    \foreach \y [count=\n,evaluate={\m=int(10-\n)}] in {4,...,-4}
    {\Vertex[x=10, y=\y, L=B_\n]{l3\m}
    \foreach \k in {1,...,9}
    {\Edges(l2\k, l3\m)}
    }
\end{tikzpicture}
}
\newcommand*\circled[1]{\tikz[baseline=(char.base)]{
            \node[shape=circle,draw,inner sep=1.8pt] (char) {#1};}}

\usepackage{skak}
\usepackage{mdframed} 
\usepackage{tkz-berge}
\newmdtheoremenv{theo}{Theorem}
\newmdtheoremenv{theox}{Definition}
\usetikzlibrary{graphs}
\usepackage{pgfplots}

\begin{document}
\maketitle
\flushbottom







\section{Introduction}

Computational arrays are ubiquitous throughout science, they are used for data-storage, and more recently for data-processing as well. In many instances the data to-be-stored or processes has the characteristic of \textit{sparsity}. From a computational perspective, this means many entries are identically zero. The use of sparse-matrices has been used for a while, but only recently with the advent of machine-learning with applications such as \cite{tensorflow} and \cite{PyTorch}, has the \textit{sparse-tensor} been defined. However, noticeably, the partial-tracing algorithms used in so many computational studies are not extended to this data structure: sparse-tensor/sparse-array. In this paper we remedy this absence of capability. This is at least one theoretically, as the implemented is on a high-level-computing-language python, and may be accessed via the python-index via the command-line command: \texttt{pip install pyftt} (ftt is a tentative name: \textit{fast-tensor-trace}, \textit{fast} because it scales linearly on the number of sparse elements, the name is playing off the fft-acronym). Results demonstrates an implementation which may outrun \cite{numpy}'s \texttt{einsum} for sparse-arrays, whose performance is directly-related it the aforementioned array's sparsity.

\section{Canonical Format \label{canonical}}

\subsection{well-ordering}

Sparse-arrays can only faithfully represent dense-arrays (bijectively) iff their row-wise elements of the index-array are unique.
In order to verify that this is the case for an arbitrarily defined index-array, they are put canonically sorted into lexicographic-order (refer to appendix \S\ref{lexorder}) to ensure they are in fact \textit{well-ordered}.
\textit{Well-ordering} is a strictly ordered/sorted list, \text{i.e.} using the relation $<$.
Such that for-all entries $a, b \in A$ ($A$ being a list/array) with element indices $i$ and $j$ respectively, then $a[i] < b[j]$ for $i<j$.
A weaker statement is \textit{partial-well-order}, this indicates the possibly for duplicates, while retaining lexicographic-order, \textit{i.e.} $\le$.
We will refer to \textit{well-ordering} an array, when discussing summation over duplicate tuple elements, while retaining lexicographic-order.



\subsection{Composition, Sub-indexing, and Permutations}

Given two 1-dimensional arrays $A$ and $B$, we can compose/subindex one by the other. Mathematically, this is written over functions with $A\circ B$, alternatively we may use a square-bracket notation, e.g. $A[B]$. If $B$ can be bijectively mapped to $A$, all entries in $B$ are unique, then this is called a \textit{permutation}. Often to sort arrays, say $A$, we determine the permutation array, $B$, such that when composed, $A[B]$, the result is ordered. This is useful when talking about sparse-arrays, because of their interconnected array structure.

\subsection{array labeling \label{labelz}}

We may associate a character/letter to every axis of a tensor/array. For instance, without-loss-of-generality if $A$ is a 4-dimensional array, then $A[i,j,k,\ell]$ denotes a labeling of $A$, for labels \{$i,j,k,\ell$\}. These characters are completely arbitrary, and are merely used to match certain indices/axës to others to create edges (or hyperedges more generally, such that every edge is associated to a type of character). 


\subsection{cartesian-product}

The \textit{cartesian-product} is all the possible 2-tuples created by all entries in $A$ (1st tuple entry), and all the entries in $B$ (2nd tuple entry). 
The cartesian-product may be generalized to products of $n$ lists-of-tuples to obtain a list of $n$-tuples of tuples. The tuple-of-tuples may be straightforwardly reduced to a larger order $m$-tuple, whereby $m>n$. Further, details of this operation are given in \S5 of \cite{Munkres}.
Originally, an operation in set-theory, when applied to lists/arrays they maintain \textit{wellorderedness} iff both original arrays were well-ordered.
If we consider the cartesian-product on partially-ordered arrays. Their cartesian-product is neither well-ordered nor partially-ordered. Naïvely we would hope that the product could produce a partially-ordered list-of-tuples. There is a permutation correction without having to resort to a naïve sort.

\subsection{direct-product}

\subsubsection{dense arrays/tensors}

Perhaps the most general operation on arrays is the direct-product (\textit{a.k.a} tensor or Kronecker-product), it is the product of all entries in one array to the other. The result is a new tensor/array with a rank equal to the sum of the constituent ranks. An example of this operation over two 2-dimensional arrays:
\begin{align*}
    A_{\mu\nu} &= 
    \begin{pmatrix}
    A_{00} & A_{01} & \cdots & A_{0n} \\
    A_{10} & A_{11} & \cdots & A_{1n} \\
    \vdots & \vdots & \ddots & \vdots \\
    A_{m0} & A_{m1} & \cdots & A_{mn} \\
    \end{pmatrix} &
    B_{\kappa\lambda} &= 
    \begin{pmatrix}
    B_{00} & B_{01} & \cdots & B_{0\ell} \\
    B_{10} & B_{11} & \cdots & B_{1\ell} \\
    \vdots & \vdots & \ddots & \vdots \\
    B_{k0} & B_{k1} & \cdots & B_{k\ell} \\
    \end{pmatrix} \quad .
\end{align*}
Then the direct-product, $\otimes$, is explicitly defined as:
\begin{align*}
    A_{\mu\nu} \otimes B_{\kappa\lambda} &= 
    \begin{pmatrix}
    A_{00} \begin{pmatrix}
    B_{00} & B_{01} & \cdots & B_{0\ell} \\
    B_{10} & B_{11} & \cdots & B_{1\ell} \\
    \vdots & \vdots & \ddots & \vdots \\
    B_{k0} & B_{k1} & \cdots & B_{k\ell} \\
    \end{pmatrix} & \cdots & \cdots & A_{0n} \begin{pmatrix}
    B_{00} & B_{01} & \cdots & B_{0\ell} \\
    B_{10} & B_{11} & \cdots & B_{1\ell} \\
    \vdots & \vdots & \ddots & \vdots \\
    B_{k0} & B_{k1} & \cdots & B_{k\ell} \\
    \end{pmatrix} \\
    \vdots & \ddots & \ddots & \vdots \\
    \vdots & \ddots & \ddots & \vdots \\
    A_{m0} \begin{pmatrix}
    B_{00} & B_{01} & \cdots & B_{0\ell} \\
    B_{10} & B_{11} & \cdots & B_{1\ell} \\
    \vdots & \vdots & \ddots & \vdots \\
    B_{k0} & B_{k1} & \cdots & B_{k\ell} \\
    \end{pmatrix} & \cdots & \cdots & A_{mn} \begin{pmatrix}
    B_{00} & B_{01} & \cdots & B_{0\ell} \\
    B_{10} & B_{11} & \cdots & B_{1\ell} \\
    \vdots & \vdots & \ddots & \vdots \\
    B_{k0} & B_{k1} & \cdots & B_{k\ell} \\
    \end{pmatrix} \\
    \end{pmatrix} \quad\quad.
\end{align*}
Symbolically this may be represented by:
\begin{align*}
    A\left[ \mu, \nu \right] \otimes B\left[ \kappa, \lambda \right] &= AB \left[ \mu, \nu , \kappa, \lambda \right]\quad\quad .
\end{align*}
The general direct-product over $n$-dimensional arrays is similarly:
\begin{align*}
    A^{(0)}\left[m_0\right] \otimes A^{(1)}\left[m_1\right] \otimes \cdots \otimes A^{(n)}\left[m_n\right] &= C\left[ m_0, m_1, \cdots, m_n \right] \quad\quad.
\end{align*}
This operation may be represented by a network/graph. 
If every data-entry in $A$ is given a node, and likewise for $B$. 
The products are associated to edges connecting them, then the complete-bipartite graph captures this product. 
For the general direct-product involving $n$ tensors/arrays the resulting graph is the complete-$n$-partite graph.
\begin{align*}
    \begin{pmatrix}
        A_1 & A_2 & A_3 \\
        A_4 & A_5 & A_6 \\
        A_7 & A_8 & A_9
    \end{pmatrix} \otimes
    \begin{pmatrix}
        B_1 & B_2 & B_3 \\
        B_4 & B_5 & B_6 \\
        B_7 & B_8 & B_9
    \end{pmatrix} &\quad\Rightarrow\quad \raisebox{-23.5ex}{\bipartnn} \quad\quad.
\end{align*}

\subsubsection{sparse arrays/tensors}

Suppose we have two well-ordered $A$ \& $B$ sparse-arrays, then their direct-product is: the binary-cartesian-product over their respective list-of-tuples, with a reshaped/flattened direct-product of their data, and concatenated shape-tuple:
\begin{align*}
    A^\text{index} \times B^\text{index} &= C^\text{index} \\
    \text{reshape} \left( A^\text{data} \otimes B^\text{data} \right) &= C^\text{data} \\
    A^\text{shape} \,\,\big|\,\big|\,\, B^\text{shape} &= C^\text{shape}\quad\quad.
\end{align*}
This readily generalizes to $n$-sparse-arrays, $A^{(n)}$:
\begin{align*}
    A^{(0)}_\text{index} \times A^{(1)}_\text{index} \times \cdots \times A^{(n)}_\text{index} &= C_\text{index} \\
    \text{reshape} \left( A^{(0)}_\text{data} \otimes A^{(1)}_\text{data} \otimes \cdots \otimes A^{(n)}_\text{data} \right) &= C_\text{data} \\
    A^{(0)}_\text{shape} \,\,\big|\,\big|\,\, A^{(1)}_\text{shape} \,\,\big|\,\big|\,\, \cdots \,\,\big|\,\big|\,\, A^{(n)}_\text{shape} &= C_\text{shape}\quad\quad.
\end{align*}

\subsection{intra-intersection/slice \label{intraintersect}}

Now suppose we are given a single tensor/array, and we would like to identify some columns/axës with each-other (via labels mentioned in \S\ref{labelz}). This reduction in the array's degrees-of-freedom is called slicing, or intra-intersecting. Because these are internal operations, these are best done first, as intra-intersecting can only reduce the size.
Unfortunately, if the intersecting columns are not adjacent to each other or the intra-intersecting indices are internal, they must be re-well-ordered (at least from the first intra-intersection column/index/axis to the rest). This operation requires an element-wise search of all rows in a given sparse-array's index-array.

\section{Reshaping}

In this section we would like to reshape sparse-arrays, as is commonly done for dense-arrays.
\textit{Reshaping} is the merging or partitioning of axes for a given array.
Unlike dense-arrays, which store data contiguously in real-space memory, sparse-arrays store the data into 3-dense-arrays. Two of these dense-arrays: shape and array-of-tuples, control the structure of the data, here we introduce dense-operations on these two arrays to alter the structure of sparse-data.
The structure of sparse-data is managed by arrays-of-tuples.
For our discussion, let $A$ and $B$ be arrays-of-tuples: $A,B \subset \prod_i \Z_i$ (\textit{i.e.} Cartesian-product of many integer array-subsets of $\Z$).
Thus we would like to define a function $h$, that maps an array-of-tuples to another array-of-tuples:
\begin{align*}
    h &: A_{} \longrightarrow B_{}\quad\quad.
\end{align*}
We choose to take an intermediate step, and instead solve the simpler problems of mapping tuples-to-integers (\S\ref{tti}) and integers-to-tuples (\S\ref{itt}):
\begin{align*}
    A_{} \longrightarrow \Z_N \longrightarrow B_{} \quad\quad.
\end{align*}

Ultimately, we desire a finite-paring functions, let $\Z_N \subset \N$ such that $|\Z_N| \ne \infty$ (\textit{i.e.} $\Z_N$ has finitely many elements), then the pairing-function is:
\begin{align*}
    f &: A_{} \longrightarrow \Z_N \\
    g &: \Z_N \longrightarrow A \quad\quad,
\end{align*}
such that: $f(g(\Z_N)) = \Z_N$ or $g(f(A)) = A$.
With the total number of dense-elements being: $N = \prod_i s[i]$. And the range of integers (in order): $
    \Z_N = \begin{pmatrix}
        0 & 1 & 2 & \cdots & N-1
    \end{pmatrix}$. 
Additionally, it is worth noting that the reshape-function defined here does not rely on any kind of ordering.
There are many kinds of pairing-functions, we choose the \textit{C-language} reshaping-order
\footnote{Let $i$ denote rows and $j$ denote columns of $A$ then:
\begin{align*}
    A_{ij} = A[i,j] &=
    \begin{pmatrix}
        1 & 2 & 3 & 4 \\
        5 & 6 & 7 & 8
    \end{pmatrix} \mathop{\xrightarrow{\hspace*{1.5cm}}}^{\text{``C''-order}}
    \begin{pmatrix}
        1 & 2 & 3 & 4 &
        5 & 6 & 7 & 8
    \end{pmatrix} = A[i'] \quad\quad.
\end{align*}}.

\subsection{modified shape \label{modshape}}

For both tuples-to-integer and integer-to-tuple maps, we need to modify the shape array, $s[i]$, into an auxiliary array $\tilde{s}[i]$.
In each case the shape-array means: current-array shape xor desired-array-shape respectively. This involves permuting the shape array in a particular fashion, with a cumulative-product. 
The modification of $s[i]$ is the following\footnote{In practice, this may be done with: 
\begin{align*}
    s[0] &= 1 \\
    s[i] &= \text{flip}(s[i]) & & (s_0, s_1, \cdots, s_{n-1}, s_n) \rightarrow (s_n, s_{n-1}, \cdots, s_{1}, s_0) \\
    s[i] &= \text{roll}(s[i], 1) & &\text{permute, via right-shift} \\
    s[i] &= \text{cumulative-product}(s[i]) \\
    s[i] &= \text{flip}(s[i]) \quad\quad.
\end{align*}}: 
\begin{align*}
    s_i &= \begin{pmatrix}
        s_0, s_1, s_2, s_3, \cdots, s_{n-1} & s_n
    \end{pmatrix} \\
    \tilde{s}_i &=
    \begin{pmatrix}
        \prod_{i\in[1,n]} s_i  & \prod_{i\in[2,n]} s_i & \prod_{i\in[3,n]} s_i  & \cdots & {s}_{n-1}s_n & s_n & 1 
    \end{pmatrix}
\end{align*}
Injectivity is therefore given by the fact that this auxiliary-shape array is strictly decreasing, \textit{i.e.} $\tilde{s}[i] > \tilde{s}[j]$ for $j > i$. And surjectivity stems from the precise spacing of the auxiliary-array. Suppose we have: $a_i = \begin{pmatrix}
        a_0 & a_1 & a_2 & \cdots & a_n
    \end{pmatrix}$, with each $a_i$ value having a ceiling value of $s_i$:
\begin{align*}
    z_i &= a_0 \left( \prod_{i\in[1,n]} s_i \right) + a_1 \left( \prod_{i\in[2,n]} s_i \right) + a_2 \left( \prod_{i\in[3,n]} s_i \right) + \cdots + a_{n-1} s_n + a_n \quad\quad.
\end{align*}

\subsection{tuples-to-integers \label{tti}}

The conversion tuples-to-integers may be achieved by a dense-matrix-vector product, for a suitably modified shape ($s_i$), array-of-tuples $A^{i}_{I}$ (with $i$ the tuple-index), and array-of-integers $a_I$:
\begin{align*}
    a_I &= A^{i}_{I} \tilde{s}_{i} \quad\quad.
\end{align*}
This operation thus has time-complexity of $\mathcal{O} \sim nN $, with $n$ being the number of columns (axës), and $N$ being the number of sparse-elements, this will have linear-scaling as usually $n <\!\!< N$.



\subsection{integers-to-tuples \label{itt}}

Given the modified shape-array, see \S\ref{modshape}, we may do the following dense-operations to obtain the new array-of-tuples, from an array-of-integers (or 1-tuples):
\begin{align*}
    A[I,i]   &= A^{I} \otimes \mathbf{1}^{i}   \\
    A[:,1:]  &= A[:,1:] \mod \tilde{s}[:-1] \\
    A[I,i] &= \bigg\lfloor \frac{A[I,i]}{\tilde{s}[i]} \bigg\rfloor \\
    A[i,I] &= A[I,i] \quad\quad.
\end{align*}
These include, computing the integer direct-product of an array-of-integers with an array consisting of all ones, $\mathbf{1}^{i}$, with the size of the desired tuple. The result is $A^{Ii}$ a 2-dimensional array.
Then the modulo-division (remainder) is applied along the second axis with the modified-shape array (along a subset of entries along that axis, \textit{i.e.} the first/zeroth entry of A[:,0] is ignored as is the last shape element).
Then the complementary-function, integer-division, is applied along the same axis, but this time over all elements.
Finally the resulting array is transposed to obtain the array-of-tuples into a \textit{canonical} form (axës following the sparse-array data entry index), hence $A[i,I]$.
This operation like, \S\ref{tti}, scales linearly in time: $\mathcal{O}\sim nN + nN + nN + 1 \sim 3nN$.


\subsection{tuples-to-tuples}

It can be easily shown that to map a tuple into another tuple of different size, using the intermediate integer is a bijective map.
This is because of the well-known relation that the composition of bijective functions is itself bijective.

\subsection{reshaping}

Using the techniques introduced in this section, any sparse-array can be cast into a sparse-matrix. Therefore we can apply sparse-matrix algorithms to sparse-arrays.

\section{direct-intersection \label{directintersect}}

We would now like to realize the inter-intersection, the intersection of elements between arrays/tensors.
The direct-intersection is the direct-product over the intersecting elements of two (or more) tensors/arrays, and deletion of non-intersecting elements. Or a particular slice/intra-intersection of their full direct-product. 
Visually, for two arrays this is given by a disconnected-union of many complete-bipartite-graphs, each complete-bipartite-graphs acting on the intersecting elements.

\begin{figure}[h]
\begin{center}
\includegraphics[width=0.90\textwidth]{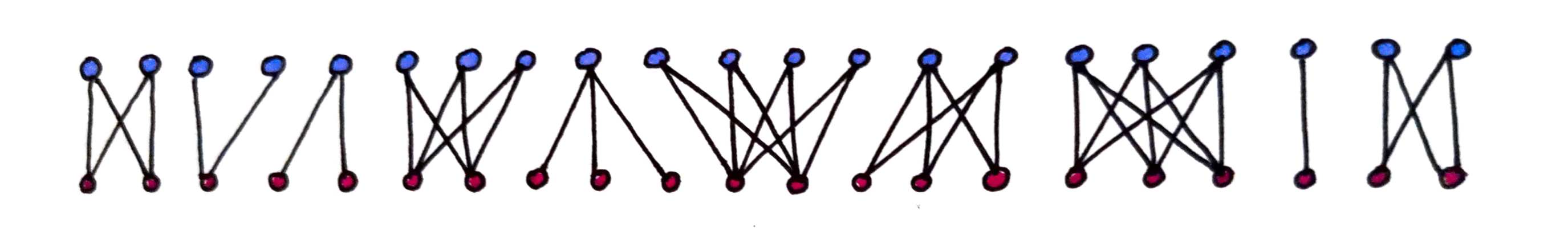}
\end{center}
\caption{ This figure shows an example of a direct-intersection (the collection of all the edges) between two arrays' (red and blue) elements (\textit{i.e.} the nodes), represented by a sum of disjoint complete bipartite-graphs. \label{bidirect} }
\end{figure}

In set-theory, the operation to determine a subset which belongs to both sets is called the intersection.
Let $A$ and $B$ be sets, then $A\cap B = C \subseteq A \,\&\, B$. 
The intersection is desirable, in our discussion, because inter-array partial-tracing only occurs over internal entries which agree in numerical value.
For arrays (sets with some arbitrary order, with potentially duplicate entries), we desire instead the indices (from each array) which contribute to the intersecting element, we desire the \textit{argument-intersection}. That is, given two arrays $A$ and $B$ and 1 intersecting element $C$ between them, $I_A$ and $I_B$ indicate the indices appearance of this element, $C$, in both $A$ and $B$. Therefore the composition/sub-index into $A$ and $B$:
\begin{align*}
    A[I_A] = B[I_B] = C \quad\quad.
\end{align*}
Now if $A$ and $B$ have multiple intersecting elements, $C^{0}, C^{1}, \cdots, C^{n}$, then we require a list-of-indices
$[I^{(0)}_A, I^{(1)}_A, \cdots, I^{(n)}_A]$, and $[I^{(0)}_B, I^{(1)}_B, \cdots, I^{(n)}_B]$ for $B$'s indices, such that $A[I_A^{(j)}] = B[I_B^{(j)}] = C^{(j)}$.
If there exists duplicates for each intersecting elements, then each $I^{(i)}_A$ are list-of-indices themselves.
Unlike the entries the list-of-indices is unique, and therefore if well-ordered their Cartesian-product will also be well-ordered. Leading to the net result of the direct-intersection, for 2 arrays, all the pairs (2-tuples) of indices that agree.
As arrays have no restriction on duplicated elements, the natural extension for the set-intersection, to arrays is the direct-intersection.


So far we have been considering two \textit{unlabeled} arrays (of the same width). If $A$ and $B$, have labels $\ell_A$ and $\ell_B$ respectively, then the label-intersection $\ell_A \cap \ell_B = e$, yields an \textit{edge}. This edge determines the columns of $A$ and $B$ which get directly-intersected.
The general-result over $n$-arrays is to obtain another sparse-array, whereby the columns are the enumeration of the arrays involved, and rows denote the row's index in each corresponding array.
\begin{align*}
    \mathop{\circled{$\bigcap$}}_{\text{edges}} A^{(n)}
\end{align*}
The result is all the possible products which contribute to the partial-trace. 




\subsection{shoelace}

Suppose we have 2 arrays $A$ and $B$, and we would like to compute their direct-intersection. Then we shall need to sort each $A$ and $B$, in order to utilize the binary-search on both arrays.
We begin by starting at one of the arrays, and searching for the first element in the other via a left-side and right-side binary-search. If found, the left-side and right-side search yields the range of values that match, else if not found the left-side and right-side search matches in value and is known as the \textit{insertion-point}.
This interval-search can be done by going between the two arrays. If we assume both arrays are sorted, then if we binary-search between the arrays, and a found element that skips over many, all these elements are by definition not included in both arrays.
This even applies if the searched element is not found, as the insertion point provides another constraint to future searches. 
Furthermore, these arrays need-not have unique elements, but merely sorted.
Additionally, all required binary-searches become increasingly constrained.
Therefore if their are $M$ unique entries in the intersection of $A$ and $B$, $|A\cap B| = M$, with $N$ total entries in each the algorithm should have time-complexity of $\mathcal{O}\sim M\log{N}$.



\subsection{multi-intersection (hyperedge)}

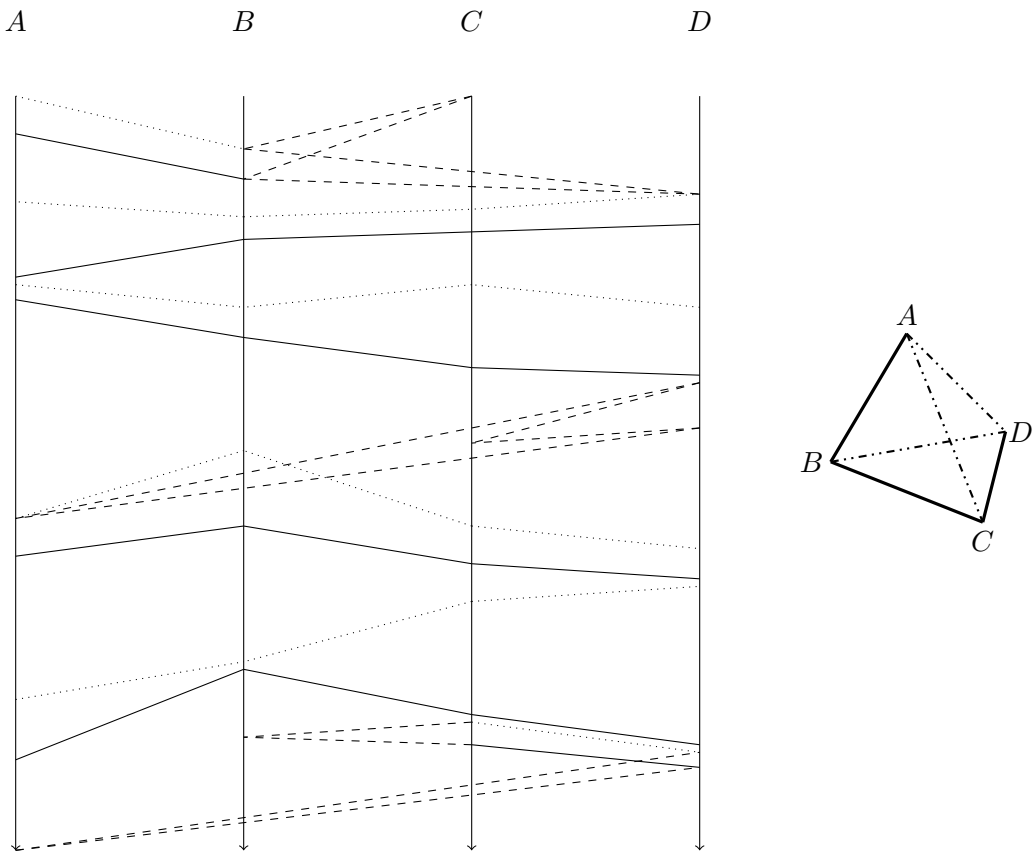
\begin{figure}[h!]
\begin{center}
\begin{tikzpicture}
    \draw[->] (0,0) -- (0,-10);
    \draw[->] (3,0) -- (3,-10);
    \draw[->] (6,0) -- (6,-10);
    \draw[->] (9,0) -- (9,-10);
    
    \draw ( 0, -0.5) -- ( 3, -1.1);
    \draw[dotted] ( 0,  0.00) -- ( 3, -0.70);
    \draw[dashed] ( 6,  0.00) -- ( 3, -0.70);
    \draw[dashed] ( 6,  0.00) -- ( 3, -1.1);
    \draw[dashed] ( 3, -0.70) -- ( 9, -1.3);
    \draw[dashed] ( 3, -1.1) -- ( 9, -1.3);

    \draw[dotted] ( 9, -1.3) -- ( 6, -1.5);
    \draw[dotted] ( 3, -1.6) -- ( 6, -1.5);
    \draw[dotted] ( 3, -1.6) -- ( 0, -1.4); 
    \draw ( 9, -1.7) -- ( 6, -1.8);
    \draw ( 3, -1.9) -- ( 6, -1.8);
    \draw ( 3, -1.9) -- ( 0, -2.4);
    
    \draw[dotted]  ( 0, -2.5) -- ( 3, -2.8);
    \draw[dotted]  ( 6, -2.5) -- ( 3, -2.8);
    \draw[dotted]  ( 9, -2.8) -- ( 6, -2.5);
    \draw ( 0, -2.7) -- ( 3, -3.2);
    \draw ( 6, -3.6) -- ( 3, -3.2);
    \draw ( 9, -3.7) -- ( 6, -3.6);
    
    \draw[dashed] ( 9, -3.8) -- ( 6, -4.6);
    \draw[dashed] ( 9, -4.4) -- ( 6, -4.6);
    \draw[dashed] ( 9, -3.8) -- ( 0, -5.6);
    \draw[dashed] ( 9, -4.4) -- ( 0, -5.6);
    
    \draw[dotted] ( 0, -5.6) -- ( 3, -4.7);
    \draw[dotted] ( 6, -5.7) -- ( 3, -4.7);
    \draw[dotted] ( 6, -5.7) -- ( 9, -6.0);
    \draw ( 0, -6.1) -- ( 3, -5.7);
    \draw ( 6, -6.2) -- ( 3, -5.7);
    \draw ( 6, -6.2) -- ( 9, -6.4);
    
    \draw[dotted] ( 6, -6.7) -- ( 9, -6.5);
    \draw[dotted] ( 6, -6.7) -- ( 3, -7.5);
    \draw[dotted] ( 0, -8.0) -- ( 3, -7.5);
    \draw ( 0, -8.8) -- ( 3, -7.6);
    \draw ( 6, -8.2) -- ( 3, -7.6);
    \draw ( 6, -8.2) -- ( 9, -8.6);
    
    \draw[dashed] ( 0, -10.0) -- ( 9, -8.7);
    \draw[dashed] ( 0, -10.0) -- ( 9, -8.9);
    \draw[dotted] ( 6, -8.3) -- ( 9, -8.7);
    \draw ( 6, -8.6) -- ( 9, -8.9);
    \draw[dashed] ( 6, -8.3) -- ( 3, -8.5);
    \draw[dashed] ( 6, -8.6) -- ( 3, -8.5);
    
    \node(draw) at ( 0, 1.0)   (a) {$A$};
    \node(draw) at ( 3, 1.0)   (a) {$B$};
    \node(draw) at ( 6, 1.0)   (a) {$C$};
    \node(draw) at ( 9, 1.0)   (a) {$D$};
\end{tikzpicture}\quad\quad
\raisebox{10em}{
\begin{tikzpicture}
    \coordinate (a) at (4,2.5);
    \coordinate (b) at (3,.8);
    \coordinate (c) at (5,0);
    \coordinate (d) at (5.3,1.2);

    \draw[very thick] (a) -- (b);
    \draw[very thick] (b) -- (c);
    \draw[very thick] (c) -- (d);
    \draw[thick, dash dot dot] (b) -- (d);
    \draw[thick, dash dot dot] (a) -- (d);
    \draw[thick, dash dot dot] (a) -- (c);

    \node at (4,2.75)  (a) {$A$};
    \node at (2.75,.8)  (a) {$B$};
    \node at (5,-0.25)  (a) {$C$};
    \node at (5.5,1.2)  (a) {$D$};
\end{tikzpicture}}
\end{center}
\caption{\label{shoelacealgor} 
(left) An example for the lace algorithm on 4 lists/arrays. (right) showing the contraction-hyperedge and the searching path (shown to the left) shown by the solid line.}
\end{figure}

The previous algorithm applies to a 2-node edge, \textit{i.e.} between 2-arrays. This is readily generalizable to a $n$-node hyper-edge, \textit{i.e.} between $n$-arrays.
The main idea here is to layout the arrays in an arbitrary 1-dimensional\footnote{Suppose we have 4 arrays as in fig. \ref{shoelacealgor}, orders of $ABCD$, $ACDB$, $CDBA$, or any permutation should yield the same results.} order with two pivot arrays. These pivot-arrays are special as they guide the search. Suppose we start the search on an edge array $A$, for some element $a \in A$, we search with the adjacent array if found we proceed to search its adjacent array, this is repeated until the last array (an edge array, $D$). If the search makes it all the way through all arrays, then their intersecting elements are direct-producted together and saved. 
Else (if any search fails) $a\in A$ is searched in $B$ for some insertion point $i$. This insertion point defines the next search, $d = D[i]$, and the algorithm repeats going towards edge array $A$. The algorithm to linking edge-to-edge may be called a \textit{sweep}. 

The algorithm is depicted in fig. \ref{shoelacealgor}, for the case of 4 list-of-tuples/arrays. On that figure, all lines (dotted, dashed, solid) are binary-searches $\mathcal{O}\sim\log{N}$, dashed lines are searches which ``failed'', yielding the insertion point (both searches yield the same location).
After a failure, one edge is compared to the other directly, the next unique element(s) in that edge start a new search.
While the dotted \& solid lines compute the left and right search respectively, and correspond to a successful search.  
Once we reach the end of any array the algorithm is finished.


\subsection{chain of edges}

Now suppose we have $n$-arrays with many edges (or hyperedges), \textit{i.e.} a network/graph, as given by the label-intersection. 
It can be easily shown, that in order to contribute to the direct-intersection of the entire network, any given element must be connected by all the network's edges to some other elements.
There must exist a path (not necessarily Eulerian) traversing all edges. This path defines a direct-product and an element to the direct-intersection.

We may take the direct-intersection over every edge in the network, to obtain intersecting sparse-arrays. Only to direct-intersect over 
The direct-intersection may be applied successively to obtain the entire-network's direct-intersection (all elements, the $n$-tuples, being unique).

\subsubsection{Tree-Search \& Ray-Tracing}

We need a method of sequentially going through all the edges in the network, in a connected fashion, this is called a path $\gamma$.
This path begins at a special, pivot, edge. 
We can initiate a \textit{scout}-binary-search, through this path and determine if makes it all the way thru the network. 
If successively traversed the network (every search yielded at least 1 element), then the entire tree-product is computed.
Else, the next intersecting element in the pivot edge begins a new scout-binary-search. Once all the intersecting elements in the pivot edge are tried the algorithm is completed.
The tree-product is the one sided direct-product. The number of products is the largest layer in the tree.



\section{the surjective-map}

So far the direct-intersection computes the intersecting-indices associated to the ordered arrays. 
This is not quite what is desired, this is because the internal-indices, which are intersected are generally in a different order than that of the external-indices.
This is not the case, as external-indices are often the objective, but its the internal indices which we have to directly-intersect. Therefore we would like to obtain the direct-intersection according to the corresponding ordered external-indices.

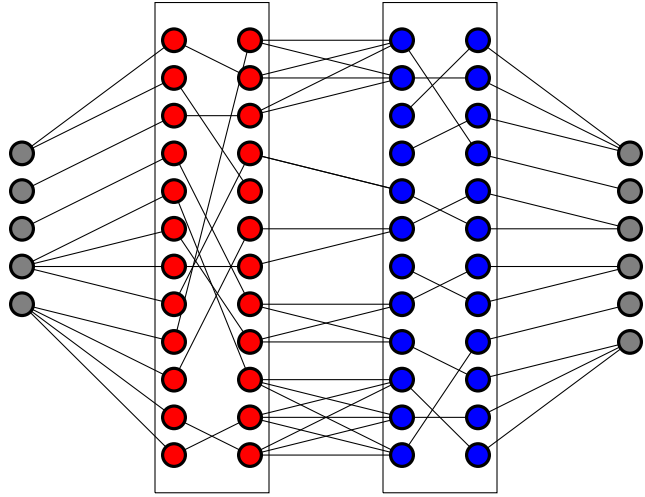
\begin{figure}[h]
\begin{center}
\begin{tikzpicture}
\draw (1.75,-0.5) -- (1.75,6.0) -- (3.25,6.0) -- (3.25,-0.5) -- (1.75,-0.5);
\draw (4.75,-0.5) -- (4.75,6.0) -- (6.25,6.0) -- (6.25,-0.5) -- (4.75,-0.5);
\draw (0,2.0) -- (2.0,0.0);
\draw (0,2.0) -- (2.0,0.5);
\draw (0,2.0) -- (2.0,1.0);
\draw (0,2.0) -- (2.0,1.5);
\draw (0,2.5) -- (2.0,2.0);
\draw (0,2.5) -- (2.0,2.5);
\draw (0,2.5) -- (2.0,3.0);
\draw (0,2.5) -- (2.0,3.5);
\draw (0,3.0) -- (2.0,4.0);
\draw (0,3.5) -- (2.0,4.5);
\draw (0,4.0) -- (2.0,5.0);
\draw (0,4.0) -- (2.0,5.5);
\draw (2.0,5.5) -- (3,5.0);
\draw (2.0,5.0) -- (3,3.5);
\draw (2.0,4.5) -- (3,4.5);
\draw (2.0,4.0) -- (3,2.0);
\draw (2.0,3.5) -- (3,1.0);
\draw (2.0,3.0) -- (3,1.5);
\draw (2.0,2.5) -- (3,2.5);
\draw (2.0,2.0) -- (3,4.0);
\draw (2.0,1.5) -- (3,5.5);
\draw (2.0,1.0) -- (3,3.0);
\draw (2.0,0.5) -- (3,0.0);
\draw (2.0,0.0) -- (3,0.5);
\draw (5.0,0.0) -- (6,1.5);
\draw (5.0,0.5) -- (6,0.5);
\draw (5.0,1.0) -- (6,0.0);
\draw (5.0,1.5) -- (6,1.0);
\draw (5.0,2.0) -- (6,2.5);
\draw (5.0,2.5) -- (6,2.0);
\draw (5.0,3.0) -- (6,3.5);
\draw (5.0,3.5) -- (6,3.0);
\draw (5.0,4.0) -- (6,4.5);
\draw (5.0,4.5) -- (6,5.5);
\draw (5.0,5.0) -- (6,5.0);
\draw (5.0,5.5) -- (6,4.0);
\draw (6.0,0.0) -- (8,1.5);
\draw (6.0,0.5) -- (8,1.5);
\draw (6.0,1.0) -- (8,1.5);
\draw (6.0,1.5) -- (8,2.0);
\draw (6.0,2.0) -- (8,2.5);
\draw (6.0,2.5) -- (8,2.5);
\draw (6.0,3.0) -- (8,3.0);
\draw (6.0,3.5) -- (8,3.0);
\draw (6.0,4.0) -- (8,3.5);
\draw (6.0,4.5) -- (8,4.0);
\draw (6.0,5.0) -- (8,4.0);
\draw (6.0,5.5) -- (8,4.0);
\draw (5.0,0.0) -- (3,0.0);
\draw (5.0,0.0) -- (3,0.5);
\draw (5.0,0.0) -- (3,1.0);
\draw (5.0,0.5) -- (3,0.0);
\draw (5.0,0.5) -- (3,0.5);
\draw (5.0,0.5) -- (3,1.0);
\draw (5.0,1.0) -- (3,0.0);
\draw (5.0,1.0) -- (3,0.5);
\draw (5.0,1.0) -- (3,1.0);
\draw (5.0,1.5) -- (3,1.5);
\draw (5.0,1.5) -- (3,2.0);
\draw (5.0,2.0) -- (3,1.5);
\draw (5.0,2.0) -- (3,2.0);
\draw (5.0,3.0) -- (3,2.5);
\draw (5.0,3.0) -- (3,3.0);
\draw (5.0,3.5) -- (3,4.0);
\draw (5.0,3.5) -- (3,4.0);
\draw (5.0,5.0) -- (3,4.5);
\draw (5.0,5.0) -- (3,5.0);
\draw (5.0,5.0) -- (3,5.5);
\draw (5.0,5.5) -- (3,4.5);
\draw (5.0,5.5) -- (3,5.0);
\draw (5.0,5.5) -- (3,5.5);
\filldraw[fill=gray, very thick](0,2.0) circle (0.15);
\filldraw[fill=gray, very thick](0,2.5) circle (0.15);
\filldraw[fill=gray, very thick](0,3.0) circle (0.15);
\filldraw[fill=gray, very thick](0,3.5) circle (0.15);
\filldraw[fill=gray, very thick](0,4.0) circle (0.15);
\filldraw[fill=red, very thick](2,0.0) circle (0.15);
\filldraw[fill=red, very thick](2,0.5) circle (0.15);
\filldraw[fill=red, very thick](2,1.0) circle (0.15);
\filldraw[fill=red, very thick](2,1.5) circle (0.15);
\filldraw[fill=red, very thick](2,2.0) circle (0.15);
\filldraw[fill=red, very thick](2,2.5) circle (0.15);
\filldraw[fill=red, very thick](2,3.0) circle (0.15);
\filldraw[fill=red, very thick](2,3.5) circle (0.15);
\filldraw[fill=red, very thick](2,4.0) circle (0.15);
\filldraw[fill=red, very thick](2,4.5) circle (0.15);
\filldraw[fill=red, very thick](2,5.0) circle (0.15);
\filldraw[fill=red, very thick](2,5.5) circle (0.15);
\filldraw[fill=red, very thick](3,0.0) circle (0.15);
\filldraw[fill=red, very thick](3,0.5) circle (0.15);
\filldraw[fill=red, very thick](3,1.0) circle (0.15);
\filldraw[fill=red, very thick](3,1.5) circle (0.15);
\filldraw[fill=red, very thick](3,2.0) circle (0.15);
\filldraw[fill=red, very thick](3,2.5) circle (0.15);
\filldraw[fill=red, very thick](3,3.0) circle (0.15);
\filldraw[fill=red, very thick](3,3.5) circle (0.15);
\filldraw[fill=red, very thick](3,4.0) circle (0.15);
\filldraw[fill=red, very thick](3,4.5) circle (0.15);
\filldraw[fill=red, very thick](3,5.0) circle (0.15);
\filldraw[fill=red, very thick](3,5.5) circle (0.15);
\filldraw[fill=blue, very thick](5,0.0) circle (0.15);
\filldraw[fill=blue, very thick](5,0.5) circle (0.15);
\filldraw[fill=blue, very thick](5,1.0) circle (0.15);
\filldraw[fill=blue, very thick](5,1.5) circle (0.15);
\filldraw[fill=blue, very thick](5,2.0) circle (0.15);
\filldraw[fill=blue, very thick](5,2.5) circle (0.15);
\filldraw[fill=blue, very thick](5,3.0) circle (0.15);
\filldraw[fill=blue, very thick](5,3.5) circle (0.15);
\filldraw[fill=blue, very thick](5,4.0) circle (0.15);
\filldraw[fill=blue, very thick](5,4.5) circle (0.15);
\filldraw[fill=blue, very thick](5,5.0) circle (0.15);
\filldraw[fill=blue, very thick](5,5.5) circle (0.15);
\filldraw[fill=blue, very thick](6,0.0) circle (0.15);
\filldraw[fill=blue, very thick](6,0.5) circle (0.15);
\filldraw[fill=blue, very thick](6,1.0) circle (0.15);
\filldraw[fill=blue, very thick](6,1.5) circle (0.15);
\filldraw[fill=blue, very thick](6,2.0) circle (0.15);
\filldraw[fill=blue, very thick](6,2.5) circle (0.15);
\filldraw[fill=blue, very thick](6,3.0) circle (0.15);
\filldraw[fill=blue, very thick](6,3.5) circle (0.15);
\filldraw[fill=blue, very thick](6,4.0) circle (0.15);
\filldraw[fill=blue, very thick](6,4.5) circle (0.15);
\filldraw[fill=blue, very thick](6,5.0) circle (0.15);
\filldraw[fill=blue, very thick](6,5.5) circle (0.15);
\filldraw[fill=gray, very thick](8,1.5) circle (0.15);
\filldraw[fill=gray, very thick](8,2.0) circle (0.15);
\filldraw[fill=gray, very thick](8,2.5) circle (0.15);
\filldraw[fill=gray, very thick](8,3.0) circle (0.15);
\filldraw[fill=gray, very thick](8,3.5) circle (0.15);
\filldraw[fill=gray, very thick](8,4.0) circle (0.15);
\end{tikzpicture}
\end{center}
\caption{ This figure shows the complete operation between two surjective-maps, two ordering, and a direct-intersection map. \label{surjective} }
\end{figure}

\subsubsection{dense-external map}

An important but subtle ingredient is the surjective-map between the external-indices and internal-indices.
Suppose we have sorted the internal and external-indices ($F$) together such that the internal indices ($D$) are lexicographically-ordered $D^{\le}$, $D^{\le}||F$ (let $||$ denote concatenation), in general this means the external-indices $F$ are not ordered. Because the external-indices yield our result, they must be well-ordered, and hence ordered. This may be achieved by a permutation $i$, to achieve partial-ordering. However, we must go further and determine which indices in the partially-ordered array are unique, these indices are given by a sorted-array $u$. When surjectively-composed/sub-indexed into $F^{\le}$ this yields $F^{<}$. This array $u$, cannot directly be used with elements of $D^{\le}$ which carry a different order and size. Therefore, we may generate an auxiliary array with the order of $F^{\le}$, but with entries according to the indices of the unique entry they correspond to in $F^{\le}$, e.g. without-loss-of-generality the $8$th unique element in $F^{\le}$ all have entry 8, this auxiliary array is in bijective agreement with $F^{\le}$ and is named $f^{\le}$. The inverse of the sorting permutation array $i^{-1}$ may be composed to revert this into the order of $D^{\le}$.
\begin{align*}
    D^{\le}||F \mathop{\longrightarrow}^{i} F^{\le} &\mathop{\longrightarrow}^{u} F^{<} \\
    F^{\le} &\mathop{\longrightarrow}^{1-1} f^{\le} \mathop{\longrightarrow}^{i^{-1}} f \quad\quad,
\end{align*}
note $f$ is in the original order of $F$, and hence ordered like $D^{\le}$. Therefore, $f$ provides a surjective map from every element in $D^{\le}$, to an element in $F^{<}$:
\begin{align*}
    f : D^{\le} \rightarrow F^{<}\quad\quad.
\end{align*}
This is important because, the result, given by the external-indices must be well-ordered, whilst they are apart of potentially many products. Therefore, it can be shown that, if $d^{(0)}\subset D^{(0)\le}$ and $d^{(1)}\subset D^{(1)\le}$ are ordered subsets of two different sparse-array list-of-tuples, such that their corresponding external-tuples are unique then:
\begin{align*}
    d^{(0)}[f^{(0)}] \times d^{(1)}[f^{(1)}] \mathop{\subset} F^{(0)<}\times F^{(1)<}\quad\quad,
\end{align*}
is a well-ordered subset of the result.

\section{2-array pure-sparse algorithm}

\begin{figure}[h]
\begin{center}
\includegraphics[width=0.65\textwidth]{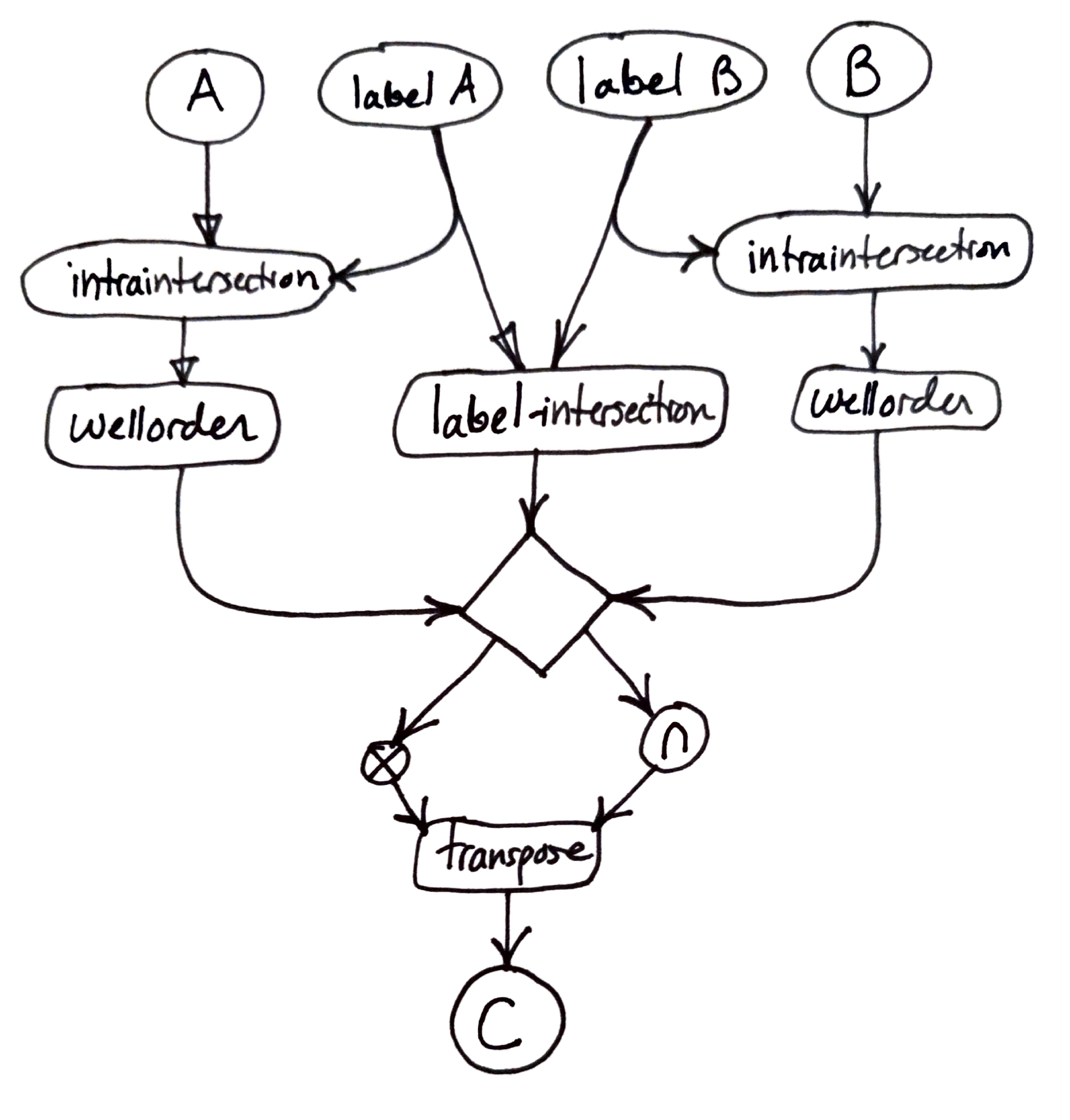}
\end{center}
\caption{ The algorithmic flow-chart for partial-tracing two sparse-arrays. \label{flowchartt} }
\end{figure}

Here we will consider the partial-trace algorithm applied for two purely sparse-arrays. This is often sufficient, as often optimum partial-tracing algorithms over many dense-arrays, involve the tracing of two arrays at a time.
Because only two arrays are considered, hyperedges (traces involving three or more arrays simultaneously) are not possible. On fig. \ref{flowchartt}. 

The direct-intersection is the crucial-piece for this algorithm, the issue is the direct-intersection typically involves a rapid expansion of elements, as it involves all the possible products involved for a partial-trace. The number of elements in the direct-intersection is also the time complexity of the algorithm (\textit{e.g.} in the naïve matrix-product, we have $\sim N^3$ products).
Therefore, we would like to perform all intra-arrays operations first, without working this the potentially large direct-intersection. This involves the intra-intersection of \S\ref{intraintersect} (removing redundant degrees-of-freedom), and establishing the canonical format of \S\ref{canonical}. This canonical format involves sorting over each internal index.


The next operation is the \textit{label-intersection}, we need to know which axës are intersected together. If no axës intersect, than the only operation is a direct-product. 
Suppose we have a sparse-array $A$, with $n$ axës, this implies the index-array, $A^\text{index}$, has $n$ columns. If we are given a label, which determines which indices are external and internal. We would like to organize the sparse-array into columns which are external xor internal. 
Multiple external-indices may be grouped together, because they do not interact with any other axës. 
While overlapping (typically internal) indices are represented by $O^{(i)}$, for each overlapping (internal) index.
with the vertical-bars indicating concatenation (over columns/axes) we have two examples of list-of-tuples/index-array decompositions:
\begin{align*}
    A^\text{index} &= F \,\big|\,\big|\, O^{(0)} \,\big|\,\big|\, O^{(1)} \,\big|\,\big|\, \cdots \,\big|\,\big|\, O^{(n-1)} \\
    B^\text{index} &= O^{(0)} \,\big|\,\big|\, O^{(1)} \,\big|\,\big|\, \cdots \,\big|\,\big|\, O^{(n)} \quad\quad.
\end{align*}
each internal-array-of-tuples corresponds to each traced/contracted edge. they are concatenated by the tuples.
This decomposition is important as the output array has at most the size of the cartesian product of the external-indices.
Continuing with the direct-intersection, we would like to determine which sparse-index get multiplied together, this is achieved with methods in \S\ref{directintersect}, by a binary-search. We then compute the direct-product over intersecting sub-array elements. For each of these elements, have associated external-indices, these are obtained via the surjective-map of \S\ref{surjective}. Like external indices are summed over to obtain the resulting array.

\section{2-array : Sparse \& Dense trace}

Here we demonstrate the partial-tracing between a purely sparse-array ($S$) and dense-array ($D$) to yield another dense-array $O$.
Let's suppose both the sparse-array ($S$) and dense-array ($D$) have both kinds of indices: external and internal.
Let's refer to the collection of external indices by $\I_E$ and $\J_E$ for both the sparse and dense-array respectively, and likewise for internal-indices $\I_I$ and $\J_I$. Such that both arrays can be brought into the form\footnote{
Due to syntax limitations, the implementation considered here involves reshaping all internal-indices for a given array into 1-axis, and all external-indices into another. Therefore the sparse-array becomes a sparse-matrix (a standard data structure, \textit{i.e.} $\I_E \rightarrow I_E$), and the dense-array becomes a dense-matrix. The same composition-procedure is then applied to this setting.
\begin{align*}
    D \mathop{\xrightarrow{\hspace*{1.5cm}}}^\text{swapaxes} D[\J_E, \J_I] &\mathop{\xrightarrow{\hspace*{1.5cm}}}^\text{reshape} D[\J_E | J_I] \\
    S \mathop{\xrightarrow{\hspace*{1.5cm}}}^\text{swapaxes} S[\I_E, \I_I] &\mathop{\xrightarrow{\hspace*{1.5cm}}}^\text{reshape} S[\I_E, I_I] \mathop{\xrightarrow{\hspace*{1.5cm}}}^\text{reshape} S[I_E,I_I] \\
    \left\{ O[\J_E, I_E] \mathop{=}^{\Sigma} S^\text{data}[I] D[\J_E, I_E]  \right\}_I \mathop{\xrightarrow{\hspace*{1.5cm}}} O \quad\quad. 
\end{align*}}: $S[\I_E\, , \I_I\,]$ and $D[\J_E\, , \J_I\,]$.
Then the resulting dense-array $O$ is $O[\I_E, \J_E]$, and it's entries are simply the \textit{iterative-sum} (sum over every sparse-entry $I$) of the product of compositions:
\begin{align*}
    O[\I_E, \J_E] &\mathop{=}^{\sum} D[\J_E\, , \hat{\I}_I\,] S^\text{data}[I]\quad\quad.
\end{align*}
Notice the sparse internal-index tuple $\I_I$ is injected/composed into the dense-array. The sub-array $D[\J_E\, , \hat{\I}_I\,]$ is multiplied by $S^\text{data}[I]$, and summed to the existing entries of $O[\I_E, \J_E]$.

\section{Results}

Here we compare the dense-contraction method to the sparse-contraction method mentioned in this paper. We will consider arrays with nonzero-entry-fractions as small as $10^{-4}$.
The algorithm is currently implemented in python, a high-level-language, and therefore suffers for optimization issues (much improvement can be done).
Despite this in certain situations it can still outrun optimized programs (optimized in lower-level-languages).
As dense-tracing-algorithms are indifferent towards the sparsity, they are calculated separately many times as a control. In the following results, the dense-partial-trace/contraction is provided by \texttt{einsum} from \cite{numpy}.
The arrays used in our results are randomly generated.
A potentially useful application of this algorithm is on tensor-networks. Therefore we consider the Matrix-Product-Operator (MPO) contraction (\S\ref{mpompo}) and a Projected-Entangled-Pair-Operators (PEPO) contraction (\S\ref{pepopepo})





\subsection{Matrix-Product}

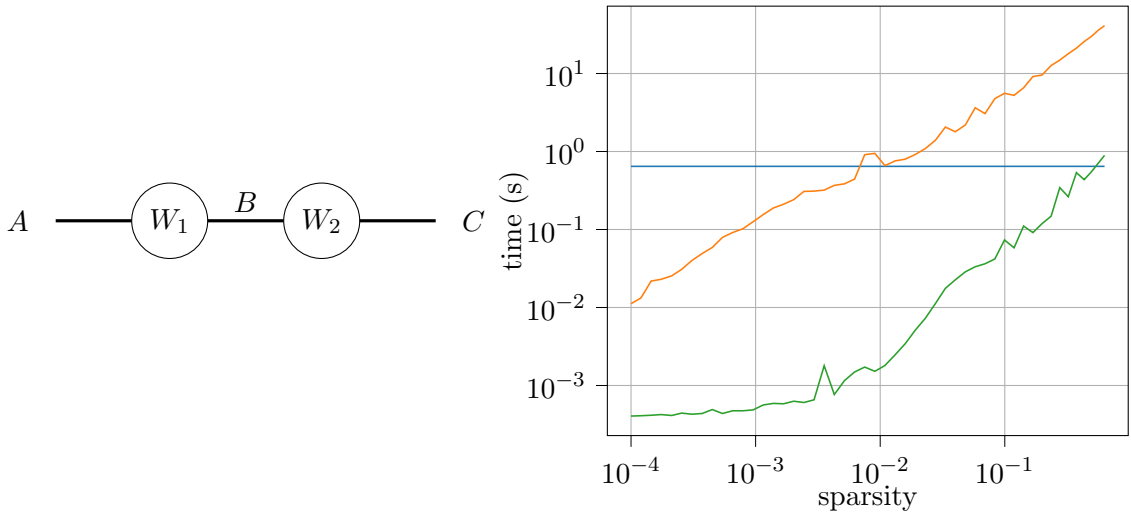
\begin{figure}[hbt!]
\begin{center}
\raisebox{9em}{
\begin{tikzpicture}
    \draw[very thick] (-0.5,0) -- (-1.5,0);
    \draw[very thick] (0.5,0) -- (1.5,0);
    \draw[very thick] (2.5,0) -- (3.5,0);
    \draw (0,0) circle [radius=0.5] node {$W_1$};
    \draw (2,0) circle [radius=0.5] node {$W_2$};

    \node at (-2.0,0)  (a) {$A$};
    \node at (1.0,0.25)  (a) {$B$};
    \node at (4.0,0)  (a) {$C$};
\end{tikzpicture}}\hspace{-2mm}
\begin{tikzpicture}

\definecolor{darkgray176}{RGB}{176,176,176}
\definecolor{darkorange25512714}{RGB}{255,127,14}
\definecolor{forestgreen4416044}{RGB}{44,160,44}
\definecolor{steelblue31119180}{RGB}{31,119,180}

\begin{axis}[
log basis x={10},
log basis y={10},
tick align=outside,
tick pos=left,
x grid style={darkgray176},
xlabel={sparsity},
xmajorgrids,
xmin=6.45595182644359e-05, xmax=0.979115887158368,
xmode=log,
xtick style={color=black},
xtick={1e-06,1e-05,0.0001,0.001,0.01,0.1,1,10},
xticklabels={
  \(\displaystyle {10^{-6}}\),
  \(\displaystyle {10^{-5}}\),
  \(\displaystyle {10^{-4}}\),
  \(\displaystyle {10^{-3}}\),
  \(\displaystyle {10^{-2}}\),
  \(\displaystyle {10^{-1}}\),
  \(\displaystyle {10^{0}}\),
  \(\displaystyle {10^{1}}\)
},
y grid style={darkgray176},
ylabel={time (s)},
ymajorgrids,
ymin=0.000227917294242681, ymax=73.0557003292253,
ymode=log,
ytick style={color=black},
ytick={1e-05,0.0001,0.001,0.01,0.1,1,10,100,1000},
yticklabels={
  \(\displaystyle {10^{-5}}\),
  \(\displaystyle {10^{-4}}\),
  \(\displaystyle {10^{-3}}\),
  \(\displaystyle {10^{-2}}\),
  \(\displaystyle {10^{-1}}\),
  \(\displaystyle {10^{0}}\),
  \(\displaystyle {10^{1}}\),
  \(\displaystyle {10^{2}}\),
  \(\displaystyle {10^{3}}\)
}
]
\addplot [semithick, steelblue31119180]
table {%
0.6321125 0.645037770271301
0.5631485 0.645037770271301
0.496802 0.645037770271301
0.4337825 0.645037770271301
0.375889 0.645037770271301
0.3233895 0.645037770271301
0.2763685 0.645037770271301
0.2354045 0.645037770271301
0.1994095 0.645037770271301
0.168151 0.645037770271301
0.1415765 0.645037770271301
0.11874 0.645037770271301
0.0994845 0.645037770271301
0.083194 0.645037770271301
0.069445 0.645037770271301
0.0579015 0.645037770271301
0.048205 0.645037770271301
0.0400815 0.645037770271301
0.0333625 0.645037770271301
0.0277255 0.645037770271301
0.0230395 0.645037770271301
0.0191305 0.645037770271301
0.0158655 0.645037770271301
0.013163 0.645037770271301
0.010924 0.645037770271301
0.0090605 0.645037770271301
0.007512 0.645037770271301
0.0062305 0.645037770271301
0.005169 0.645037770271301
0.004282 0.645037770271301
0.003552 0.645037770271301
0.0029435 0.645037770271301
0.00244 0.645037770271301
0.002022 0.645037770271301
0.001675 0.645037770271301
0.0013885 0.645037770271301
0.0011505 0.645037770271301
0.000954 0.645037770271301
0.00079 0.645037770271301
0.000655 0.645037770271301
0.000542 0.645037770271301
0.000449 0.645037770271301
0.000372 0.645037770271301
0.000308 0.645037770271301
0.000255 0.645037770271301
0.000212 0.645037770271301
0.000175 0.645037770271301
0.000145 0.645037770271301
0.00012 0.645037770271301
0.0001 0.645037770271301
};
\addplot [semithick, darkorange25512714]
table {%
0.6321125 41.0569779872894
0.5631485 35.9203531742096
0.496802 29.9557948112488
0.4337825 25.6906313896179
0.375889 21.1466858386993
0.3233895 17.9336483478546
0.2763685 14.8549842834473
0.2354045 12.6528789997101
0.1994095 9.575519323349
0.168151 9.1312563419342
0.1415765 6.55700826644897
0.11874 5.24738144874573
0.0994845 5.5753812789917
0.083194 4.75843596458435
0.069445 3.05953741073608
0.0579015 3.63428664207458
0.048205 2.18700909614563
0.0400815 1.79229712486267
0.0333625 2.05814075469971
0.0277255 1.39065480232239
0.0230395 1.09131956100464
0.0191305 0.915568113327026
0.0158655 0.797092914581299
0.013163 0.757791519165039
0.010924 0.658373355865479
0.0090605 0.946691036224365
0.007512 0.910290479660034
0.0062305 0.443202018737793
0.005169 0.385010480880737
0.004282 0.367893695831299
0.003552 0.319850921630859
0.0029435 0.310583114624023
0.00244 0.307824373245239
0.002022 0.241397142410278
0.001675 0.210672855377197
0.0013885 0.188733100891113
0.0011505 0.155169486999512
0.000954 0.125649690628052
0.00079 0.102133989334106
0.000655 0.0915391445159912
0.000542 0.0791676044464111
0.000449 0.0589146614074707
0.000372 0.0491266250610352
0.000308 0.0400278568267822
0.000255 0.0308487415313721
0.000212 0.0254495143890381
0.000175 0.0230162143707275
0.000145 0.0218174457550049
0.00012 0.0132434368133545
0.0001 0.0111782550811768
};
\addplot [semithick, forestgreen4416044]
table {%
0.6321125 0.891834497451782
0.5631485 0.709275245666504
0.496802 0.5511314868927
0.4337825 0.435818195343018
0.375889 0.53615140914917
0.3233895 0.263120412826538
0.2763685 0.344343662261963
0.2354045 0.147867918014526
0.1994095 0.119200706481934
0.168151 0.0912356376647949
0.1415765 0.11073112487793
0.11874 0.0582485198974609
0.0994845 0.0733928680419922
0.083194 0.0418562889099121
0.069445 0.0363695621490479
0.0579015 0.033275842666626
0.048205 0.0286462306976318
0.0400815 0.0226306915283203
0.0333625 0.0176076889038086
0.0277255 0.0112080574035645
0.0230395 0.00728225708007812
0.0191305 0.00511837005615234
0.0158655 0.00342631340026855
0.013163 0.0024714469909668
0.010924 0.00180673599243164
0.0090605 0.00151968002319336
0.007512 0.00172281265258789
0.0062305 0.00149130821228027
0.005169 0.00115561485290527
0.004282 0.000766515731811523
0.003552 0.00179028511047363
0.0029435 0.000654697418212891
0.00244 0.000605344772338867
0.002022 0.000628471374511719
0.001675 0.000582456588745117
0.0013885 0.000589370727539062
0.0011505 0.000562429428100586
0.000954 0.000486135482788086
0.00079 0.000473976135253906
0.000655 0.000473976135253906
0.000542 0.000437736511230469
0.000449 0.000491142272949219
0.000372 0.000437259674072266
0.000308 0.000428438186645508
0.000255 0.0004425048828125
0.000212 0.000412464141845703
0.000175 0.000423669815063477
0.000145 0.000415325164794922
0.00012 0.000409364700317383
0.0001 0.000405550003051758
};
\end{axis}

\end{tikzpicture}
\end{center}
\caption{ Partial-trace of two matrices. (left) A diagram representing the matrix-product. (right) a plot showing the sparsity time-complexity, with \texttt{einsum} in blue, our algorithm in orange, and SciPy's sparse-matrix product in green. \label{mptr} }
\end{figure}

Here we consider products between two 2-dimensional arrays $W^{(1)}_{AB}W^{(2)}_{BC}$ ($\mathcal{O}\sim N^3$), this is useful comparison because of the established theory of sparse-matrices. Because our algorithm, mentioned in this paper is not optimized in a lower-level language, its comparison to an established code might give insight on the optimized performance.
The established sparse-matrix code is provided from \cite{scipy}.
As shown by the results our-implementation is approximately $50$-fold slower than SciPy's sparse-matrix-code, for dense and very-sparse limits.
On fig. \ref{mptr}, we see the dense-sparse cross-over point, for our algorithm, is approximately 0.01, while for SciPy's implementation it is $\approx 0.50$.

\subsection{Matrix-Product-Operators Trace \label{mpompo}}

\begin{figure}[hbt!]
\begin{center}
\raisebox{4.5em}{
\begin{tikzpicture}
    \draw[very thick] (-0.5,0) -- (-1.5,0);
    \draw[very thick] (0.5,0) -- (1.5,0);
    \draw[very thick] (2.5,0) -- (3.5,0);
    \draw[very thick] (0.0,0.5) -- (0.0,1.5);
    \draw[very thick] (0.0,-0.5) -- (0.0,-1.5);
    \draw[very thick] (2.0,-0.5) -- (2.0,-1.5);
    \draw[very thick] (2.0,0.5) -- (2.0,1.5);
    \draw (0,0) circle [radius=0.5] node {$W_1$};
    \draw (2,0) circle [radius=0.5] node {$W_2$};

    \node at (0.0,2.0)  (a) {$a$};
    \node at (0.0,-2.0)  (a) {$b$};
    \node at (2.0,2.0)  (a) {$c$};
    \node at (2.0,-2.0)  (a) {$d$};
    \node at (-2.0,0)  (a) {$A$};
    \node at (1.0,0.25)  (a) {$B$};
    \node at (4.0,0)  (a) {$C$};
\end{tikzpicture}}
\hspace{-2mm}
\begin{tikzpicture}

\definecolor{darkgray176}{RGB}{176,176,176}
\definecolor{darkorange25512714}{RGB}{255,127,14}
\definecolor{steelblue31119180}{RGB}{31,119,180}

\begin{axis}[
log basis x={10},
log basis y={10},
tick align=outside,
tick pos=left,
x grid style={darkgray176},
xlabel={sparsity},
xmajorgrids,
xmin=6.45593108138682e-05, xmax=0.979181960015913,
xmode=log,
xtick style={color=black},
xtick={1e-06,1e-05,0.0001,0.001,0.01,0.1,1,10},
xticklabels={
  \(\displaystyle {10^{-6}}\),
  \(\displaystyle {10^{-5}}\),
  \(\displaystyle {10^{-4}}\),
  \(\displaystyle {10^{-3}}\),
  \(\displaystyle {10^{-2}}\),
  \(\displaystyle {10^{-1}}\),
  \(\displaystyle {10^{0}}\),
  \(\displaystyle {10^{1}}\)
},
y grid style={darkgray176},
ylabel={time (s)},
ymajorgrids,
ymin=0.00161245452104196, ymax=63.2236198143801,
ymode=log,
ytick style={color=black},
ytick={0.0001,0.001,0.01,0.1,1,10,100,1000},
yticklabels={
  \(\displaystyle {10^{-4}}\),
  \(\displaystyle {10^{-3}}\),
  \(\displaystyle {10^{-2}}\),
  \(\displaystyle {10^{-1}}\),
  \(\displaystyle {10^{0}}\),
  \(\displaystyle {10^{1}}\),
  \(\displaystyle {10^{2}}\),
  \(\displaystyle {10^{3}}\)
}
]
\addplot [semithick, steelblue31119180]
table {%
0.632153125 4.41068117618561
0.56306875 4.41068117618561
0.49625625 4.41068117618561
0.434415625 4.41068117618561
0.375790625 4.41068117618561
0.322446875 4.41068117618561
0.276821875 4.41068117618561
0.2353625 4.41068117618561
0.19890625 4.41068117618561
0.16804375 4.41068117618561
0.1416 4.41068117618561
0.118703125 4.41068117618561
0.099371875 4.41068117618561
0.082965625 4.41068117618561
0.06943125 4.41068117618561
0.057803125 4.41068117618561
0.048225 4.41068117618561
0.0401375 4.41068117618561
0.03339375 4.41068117618561
0.02778125 4.41068117618561
0.02299375 4.41068117618561
0.0190875 4.41068117618561
0.0158625 4.41068117618561
0.013175 4.41068117618561
0.010896875 4.41068117618561
0.009053125 4.41068117618561
0.007503125 4.41068117618561
0.006225 4.41068117618561
0.005165625 4.41068117618561
0.004284375 4.41068117618561
0.003540625 4.41068117618561
0.002934375 4.41068117618561
0.0024375 4.41068117618561
0.00201875 4.41068117618561
0.001675 4.41068117618561
0.0013875 4.41068117618561
0.00115 4.41068117618561
0.00095 4.41068117618561
0.00078125 4.41068117618561
0.000646875 4.41068117618561
0.0005375 4.41068117618561
0.00044375 4.41068117618561
0.00036875 4.41068117618561
0.000303125 4.41068117618561
0.00025 4.41068117618561
0.00020625 4.41068117618561
0.000175 4.41068117618561
0.00014375 4.41068117618561
0.00011875 4.41068117618561
0.0001 4.41068117618561
};
\addplot [semithick, darkorange25512714]
table {%
0.632153125 39.0920834541321
0.56306875 27.0296814441681
0.49625625 25.0578689575195
0.434415625 22.359167098999
0.375790625 20.1442153453827
0.322446875 16.4077031612396
0.276821875 21.405636548996
0.2353625 13.4558200836182
0.19890625 11.4228115081787
0.16804375 10.4664785861969
0.1416 9.65957927703857
0.118703125 7.92048931121826
0.099371875 8.43696689605713
0.082965625 6.95191049575806
0.06943125 5.37880492210388
0.057803125 5.26477551460266
0.048225 3.73571395874023
0.0401375 2.96987867355347
0.03339375 2.83988857269287
0.02778125 2.45573592185974
0.02299375 1.45126557350159
0.0190875 1.14120841026306
0.0158625 0.897236585617065
0.013175 0.616830825805664
0.010896875 0.492286920547485
0.009053125 0.341269254684448
0.007503125 0.282267808914185
0.006225 0.189735174179077
0.005165625 0.141971588134766
0.004284375 0.13997220993042
0.003540625 0.0899732112884521
0.002934375 0.0786805152893066
0.0024375 0.0657289028167725
0.00201875 0.0448007583618164
0.001675 0.0344176292419434
0.0013875 0.0459837913513184
0.00115 0.0232570171356201
0.00095 0.0179805755615234
0.00078125 0.014859676361084
0.000646875 0.0125699043273926
0.0005375 0.0107367038726807
0.00044375 0.0105054378509521
0.00036875 0.00654411315917969
0.000303125 0.00557088851928711
0.00025 0.00464630126953125
0.00020625 0.00364494323730469
0.000175 0.00329875946044922
0.00014375 0.00263261795043945
0.00011875 0.00412988662719727
0.0001 0.00260782241821289
};
\end{axis}

\end{tikzpicture}
\end{center}
\caption{ Partial-trace of two MPO-elements. (left) the diagram showing the partial-trace. (right) a plot showing the sparsity time-complexity, with \texttt{einsum} in blue and our-algorithm in orange. \label{mpotr} }
\end{figure}
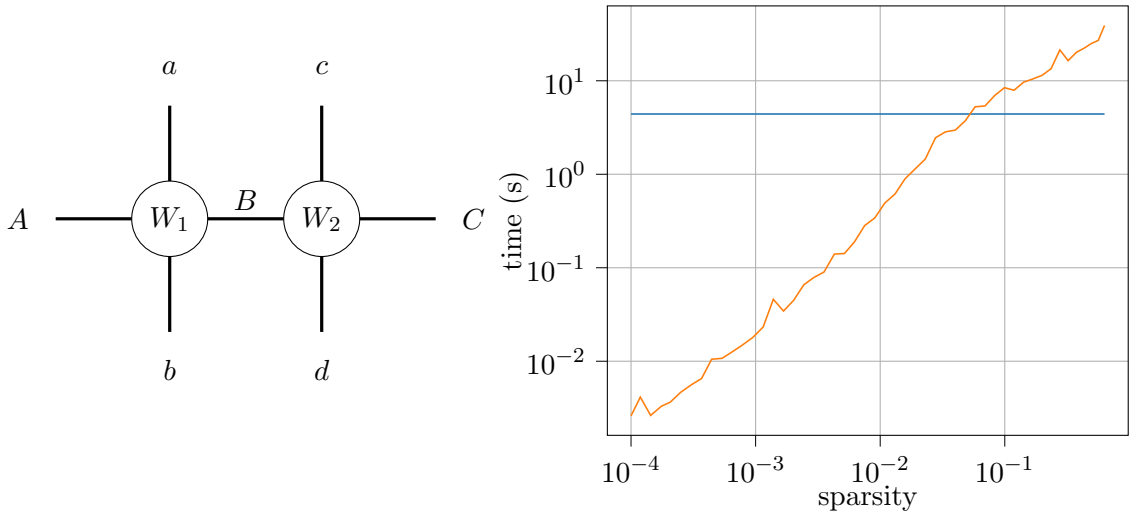

Next we consider the partial-trace of two sparse 4-index Matrix-Product-Operators (MPOs): $W^{(1)}_{ABab} W^{(2)}_{BCcd} = C_{ACabcd}$ ($\mathcal{O}\sim N^7$). In our test each 4-index operator has shape $s[i] = (20,20,20,20)$. The fig. \ref{mpotr}, shows our results. There is a dense-sparse cross-over point of approximately 0.05, and the time-complexity is shown to scale directly with sparsity.

\subsection{Projected-Entangled-Pair-Operators Trace \label{pepopepo}}

\begin{figure}[hbt!]
\begin{center}
\raisebox{5em}{
\begin{tikzpicture}
    \draw[very thick] (-0.5,0) -- (-1.5,0);
    \draw[very thick] (0.5,0) -- (1.5,0);
    \draw[very thick] (2.5,0) -- (3.5,0);
    \draw[very thick] (0.0,0.5) -- (0.0,1.5);
    \draw[very thick] (0.0,-0.5) -- (0.0,-1.5);
    \draw[very thick] (2.0,-0.5) -- (2.0,-1.5);
    \draw[very thick] (2.0,0.5) -- (2.0,1.5);
    \draw (0,0) circle [radius=0.5] node {$W_1$};
    \draw (2,0) circle [radius=0.5] node {$W_2$};

    \draw[very thick] (0.35,0.35) -- (0.75,0.75);
    \draw[very thick] (-0.35,-0.35) -- (-0.75,-0.75);
    \draw[very thick] (2.35,0.35) -- (2.75,0.75);
    \draw[very thick] (1.65,-0.35) -- (1.25,-0.75);

    \node at (0.0,2.0)  (a) {$a$};
    \node at (0.0,-2.0)  (a) {$b$};
    \node at (2.0,2.0)  (a) {$c$};
    \node at (2.0,-2.0)  (a) {$d$};
    \node at (-2.0,0)  (a) {$A$};
    \node at (1.0,1.0)  (a) {$B$};
    \node at (-1.0,-1.0)  (a) {$C$};
    \node at (1.0,0.25)  (a) {$D$};
    
    \node at (3.0,1.0)  (a) {$E$};
    \node at (1.0,-1.0)  (a) {$F$};
    \node at (4.0,0)  (a) {$G$};
\end{tikzpicture}}\hspace{-3mm}
\begin{tikzpicture}

\definecolor{darkgray176}{RGB}{176,176,176}
\definecolor{darkorange25512714}{RGB}{255,127,14}
\definecolor{steelblue31119180}{RGB}{31,119,180}

\begin{axis}[
log basis x={10},
log basis y={10},
tick align=outside,
tick pos=left,
x grid style={darkgray176},
xlabel={sparsity},
xmajorgrids,
xmin=5.32998754456739e-06, xmax=0.014240706792744,
xmode=log,
xtick style={color=black},
xtick={1e-07,1e-06,1e-05,0.0001,0.001,0.01,0.1,1},
xticklabels={
  \(\displaystyle {10^{-7}}\),
  \(\displaystyle {10^{-6}}\),
  \(\displaystyle {10^{-5}}\),
  \(\displaystyle {10^{-4}}\),
  \(\displaystyle {10^{-3}}\),
  \(\displaystyle {10^{-2}}\),
  \(\displaystyle {10^{-1}}\),
  \(\displaystyle {10^{0}}\)
},
y grid style={darkgray176},
ylabel={time (s)},
ymajorgrids,
ymin=0.000491577394963754, ymax=22.9276420956044,
ymode=log,
ytick style={color=black},
ytick={1e-05,0.0001,0.001,0.01,0.1,1,10,100,1000},
yticklabels={
  \(\displaystyle {10^{-5}}\),
  \(\displaystyle {10^{-4}}\),
  \(\displaystyle {10^{-3}}\),
  \(\displaystyle {10^{-2}}\),
  \(\displaystyle {10^{-1}}\),
  \(\displaystyle {10^{0}}\),
  \(\displaystyle {10^{1}}\),
  \(\displaystyle {10^{2}}\),
  \(\displaystyle {10^{3}}\)
}
]
\addplot [semithick, steelblue31119180]
table {%
0.00994873046875 14.0650956392288
0.008636474609375 14.0650956392288
0.00752067565917969 14.0650956392288
0.00652122497558594 14.0650956392288
0.00566291809082031 14.0650956392288
0.00492286682128906 14.0650956392288
0.00428199768066406 14.0650956392288
0.00372314453125 14.0650956392288
0.00323295593261719 14.0650956392288
0.00280570983886719 14.0650956392288
0.00243759155273438 14.0650956392288
0.00211334228515625 14.0650956392288
0.00183677673339844 14.0650956392288
0.00159645080566406 14.0650956392288
0.0013885498046875 14.0650956392288
0.0012054443359375 14.0650956392288
0.00104522705078125 14.0650956392288
0.00090789794921875 14.0650956392288
0.00078582763671875 14.0650956392288
0.000684738159179688 14.0650956392288
0.0005950927734375 14.0650956392288
0.000514984130859375 14.0650956392288
0.000446319580078125 14.0650956392288
0.00038909912109375 14.0650956392288
0.000335693359375 14.0650956392288
0.000293731689453125 14.0650956392288
0.000255584716796875 14.0650956392288
0.00022125244140625 14.0650956392288
0.00019073486328125 14.0650956392288
0.000164031982421875 14.0650956392288
0.00014495849609375 14.0650956392288
0.000125885009765625 14.0650956392288
0.0001068115234375 14.0650956392288
9.5367431640625e-05 14.0650956392288
8.0108642578125e-05 14.0650956392288
6.866455078125e-05 14.0650956392288
6.103515625e-05 14.0650956392288
5.340576171875e-05 14.0650956392288
4.57763671875e-05 14.0650956392288
3.814697265625e-05 14.0650956392288
3.4332275390625e-05 14.0650956392288
3.0517578125e-05 14.0650956392288
2.6702880859375e-05 14.0650956392288
2.288818359375e-05 14.0650956392288
1.9073486328125e-05 14.0650956392288
1.52587890625e-05 14.0650956392288
1.52587890625e-05 14.0650956392288
1.1444091796875e-05 14.0650956392288
1.1444091796875e-05 14.0650956392288
7.62939453125e-06 14.0650956392288
};
\addplot [semithick, darkorange25512714]
table {%
0.00994873046875 1.43381524085999
0.008636474609375 1.12512445449829
0.00752067565917969 0.84903621673584
0.00652122497558594 0.681174516677856
0.00566291809082031 0.542681455612183
0.00492286682128906 0.445064783096313
0.00428199768066406 0.337268114089966
0.00372314453125 0.276551961898804
0.00323295593261719 0.213735818862915
0.00280570983886719 0.157876014709473
0.00243759155273438 0.134737730026245
0.00211334228515625 0.143460273742676
0.00183677673339844 0.157458305358887
0.00159645080566406 0.136191129684448
0.0013885498046875 0.107608079910278
0.0012054443359375 0.0972418785095215
0.00104522705078125 0.0782530307769775
0.00090789794921875 0.0768826007843018
0.00078582763671875 0.0537979602813721
0.000684738159179688 0.049760103225708
0.0005950927734375 0.0370862483978271
0.000514984130859375 0.0317025184631348
0.000446319580078125 0.0260591506958008
0.00038909912109375 0.0288197994232178
0.000335693359375 0.0221850872039795
0.000293731689453125 0.0152351856231689
0.000255584716796875 0.017533540725708
0.00022125244140625 0.0119724273681641
0.00019073486328125 0.0103814601898193
0.000164031982421875 0.00894069671630859
0.00014495849609375 0.00772190093994141
0.000125885009765625 0.00768065452575684
0.0001068115234375 0.00557756423950195
9.5367431640625e-05 0.0054018497467041
8.0108642578125e-05 0.00584697723388672
6.866455078125e-05 0.00487780570983887
6.103515625e-05 0.00333642959594727
5.340576171875e-05 0.00371980667114258
4.57763671875e-05 0.00273275375366211
3.814697265625e-05 0.00269341468811035
3.4332275390625e-05 0.00232267379760742
3.0517578125e-05 0.00203466415405273
2.6702880859375e-05 0.00185537338256836
2.288818359375e-05 0.00163435935974121
1.9073486328125e-05 0.00137090682983398
1.52587890625e-05 0.00129485130310059
1.52587890625e-05 0.00130558013916016
1.1444091796875e-05 0.000947475433349609
1.1444091796875e-05 0.000990629196166992
7.62939453125e-06 0.000801324844360352
};
\end{axis}

\end{tikzpicture}
\end{center}
\caption{ Partial-trace of two PEPO-elements. (left) the diagram showing the partial-trace. (right) a plot showing the sparsity time-complexity, with \texttt{einsum} in blue and our-algorithm in orange. \label{pepotr}  }
\end{figure}
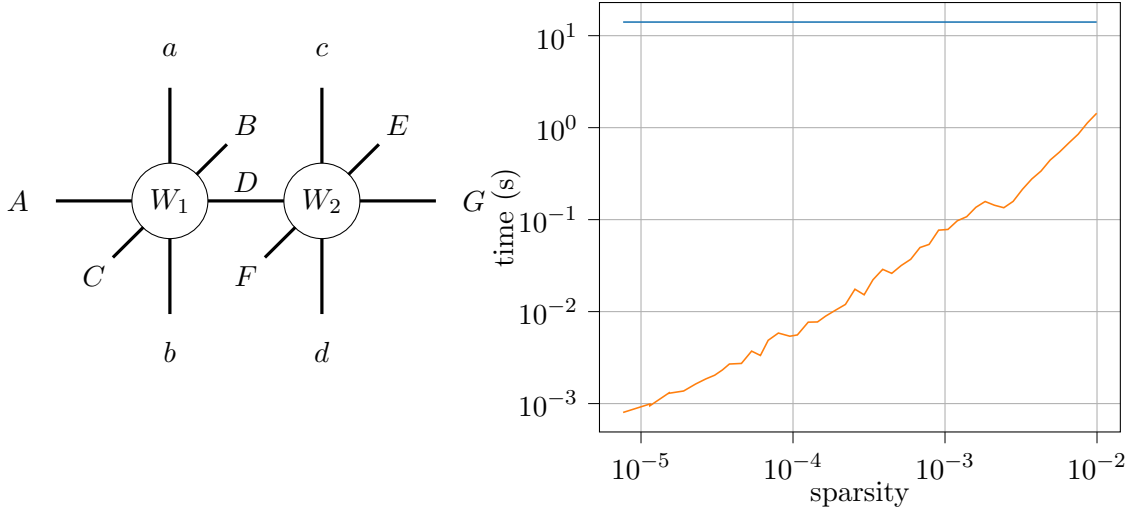

Next we consider the partial-trace of two sparse 6-index Projected-Entangled-Pair-Operator (PEPOs): $W^{(1)}_{ABCDab} W^{(2)}_{DEFGcd} = C_{ABCEFGabcd}$ ($\mathcal{O}\sim N^{11}$). In our test each 4-index operator has shape $s[i] = (8,8,8,8,8,8)$. Our results are shown in fig. \ref{pepotr}, and demonstrate that the algorithm still scales linearly with the sparsity.

\section{Summary}

As shown by our results, sparse-arrays can be used to leverage certain advantages, partial-tracing, over dense-arrays. However, there are some manipulations which are seemingly slower.
Compared to dense-array manipulations, dense-arrays algorithms can access elements in $\mathcal{O}(1)$, while for this algorithm they require $\mathcal{O}(\log N)$, and reshaping dense-arrays can be done in $\mathcal{O}(1)$, while for sparse-arrays its done in $\mathcal{O}(N)$. Additionally, sparse-arrays need to be sorted along each index leading to $\mathcal{O}(n N\log N)$ operations for a sparse-array with $n$ columns/indices and $N$ entries. The sorting overhead are additive and hence maximally scale to the number of initial elements, all which can be done in parallel.
In our results the exponential scaling relating to sparsity is approximately 1 (from 0.86, 1.10, and 1.17 respectively) for all tests considered. This implies the algorithmic time-complexity depends directly on the sparsity.
There is still more work to do, including optimizing the algorithm in a low-level-computing-language, partial-tracing by leveraging symmetries, and a full extension for block-sparse data structures.



\appendix

\section{sparse-arrays}

\subsection{attributes}

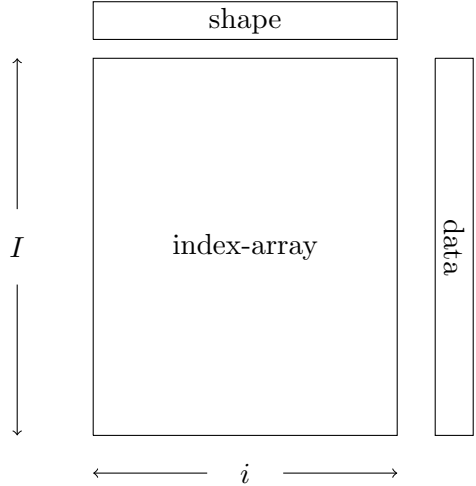
\begin{figure}[h!]
\begin{center}
\begin{tikzpicture}
    \draw[<-] (-1, 0) -- (-1, 2);
    \draw[->] (-1, 3) -- (-1, 5.0);
    \node(draw) at (-1, 2.5)  (a) {$I$};
    \draw[<-] (0, -0.5) -- (1.5, -0.5);
    \draw[<-] (4, -0.5) -- (2.5, -0.5);
    \node(draw) at (2.0, -0.5)  (a) {$i$};
    
    \draw (0,0) -- (0,5) -- (4,5) -- ( 4,0) -- (0,0);
    \draw (4.5,5) -- (5.0,5) -- ( 5.0,0) -- ( 4.5,0) -- (4.5,5);
    \draw (0,5.25) -- (0,5.75) -- (4,5.75) -- ( 4,5.25) -- (0,5.25);

    \node(draw) at (2.00, 5.50)   (a) {shape};
    \node(draw) at (4.75, 2.50)   (a) {\rotatebox{270}{data}};
    \node(draw) at (2.00, 2.50)   (a) {index-array};
    
\end{tikzpicture}
\end{center}
\caption{\label{sparsedecomp} The decomposition of the sparse-array into a shape, array-of-tuples, and data arrays.}
\end{figure}

Sparse-Arrays are meant to represent dense-arrays perfectly, with the assertion that elements which are zero are neglected\footnote{sometimes approximately zero, in which case the representation is approximate.}.
The sparse-arrays may be broken into 2 primary parts: An 2-dimensional dense index-array and 1-dimensional dense data-array . 
The sparse index-array is represented by:
\begin{align*}
    \text{Sparse Index-Array} \equiv A\left[ \quad | \quad \right]\quad\quad,
\end{align*}
with the vertical-line, $|$, separating this 2d array's two indices. 
With the left-side representing the dense-index columns, and the right-side indexing the enumeration of the sparse-tuple entry. In order to draw parallel with the dense-representation, the dense-indices are often shown explicitly included.
For example for a tensor $A$, with dense-indices $ijkl$, and sparse-index $I$ (enumerating all the non-zero values):
\begin{align*}
    \text{Dense  Representation} &\equiv A\left[ i, j, k, l \right] \\
    \text{Sparse Representation} &\equiv A\left[ i, j, k, l \,|\, I\, \right]\quad\quad.
\end{align*}
Notice, the dense representation does not have the $I$ (sparse) index.
Additionally, a small bit of information is the shape of the dense-representation of the array.
Therefore the complete attributes of the sparse-array include: shape-array ($s$), array-of-tuples/index-array ($A_\text{index}$), and the data-array ($A_\text{data}$). The full anatomy of a sparse-array is thus given in fig. \ref{sparsedecomp}.



\subsubsection{shape}

The shape-array is a 1-dimensional array which gives the shape of the array in it's dense-representation. This is useful for the mapping between sparse-arrays and dense-arrays.

\subsubsection{indices}

The sparse-array index-array or \textit{indices} or \textit{array-of-tuples} is by convention lexicographically-ordered (ssort, by convention in ascending-order). This helps in identifying duplicates, and other sparse-array operations. 
The index-array captures the structure of the array.

\subsubsection{data}

For us the \textit{data-arrays} are 1-dimensional with a length matching the index-arrays, and being in 1-1 correspondence with the index-array. That is the list-index (of the index-array) matches the data-array index. These arrays are typically of \texttt{float} datatypes, and have no restriction on duplicates, with a slot-representation of $A_\text{data}\left[\quad\right]$.

\subsection{sparsity}

The sparsity is defined by (assuming $A_\text{index}$ is in a proper format, only filled with unique entries):
\begin{align*}
    \text{sparsity} &= \frac{\text{\text{total elements in proper sparse-tensor}}}{\text{total elements in dense-tensor}}\quad\quad \\
    &= \frac{\text{len}(A_\text{data})}{\prod_i \text{shape}[i]} \quad\quad.
\end{align*}

\subsection{well-ordering \label{clean}}

Suppose we have a ssorted index-array, $A_\text{index}\left[\!\left[ \quad | \quad \right]\!\right]$, and it's uniques array, $u$ (indicating the first appearance of a unique element in the sorted array). Then the duplicate elements are all the elements in the ssorted sparse-tensor between two successive unique elements, $u[i] : u[i+1]$ (not including $u[i+1]$). Therefore, we sum all these elements of $A\left[\!\left[ \quad | \quad \right]\!\right]$ for all successive-elements in $u$. This creates an array with unique data values for each unique index-tuple:
\begin{align*}
    A_\text{index}\left[\!\left[\!\left[ \quad | \quad \right]\!\right]\!\right] &= A_\text{index}\left[\!\left[ \quad \,\,|\,\,u \,\,\right]\!\right]\quad, \\
    A_\text{data}\left[ \quad \right] &= \sum_{i\in|u|} A_\text{data}\left[\,\, u[i] : u[i+1] \,\,\right]\quad\quad.
\end{align*}
Note $u[i+1] = |u|$ for the last unique-array's index, $i = |u|$.





\section{lexicographic-order \label{lexorder}}





\subsection{array-of-tuples}


Numerical \textit{Lexicographic-order} is defined on a 2-dimensional array, a matrix (over some data-type), and we would like to sort along a given axis, say the columns, while preserving the rows. Alternatively, we may call this data-structure as a \textit{array-of-tuples}, in order to distinguish that the \textit{tuples'} order is preserved (rows), and we permute the \textit{array}, i.e. columns. We may denote this 2d array, array-of-tuples as: $A[ n | I ]$, for a \textit{array-index} (denoted by $I$) that acts \textit{row-wise} refers to the $I$th tuple, while acting \textit{column-wise} with a \textit{tuple-index} (denoted by $n$).




\subsection{tuple-tuple comparison \label{tuplecompare}}

Now we would like to define a notion of a tuple being smaller/larger than another, we denote this comparison by $\prec$ xor $\preceq$.
Suppose we have two tuples (of the same finite length, filled in with $n+1$ entries) $a, b \in \N^{n+1}$, then $a < b$ if we preform an element-wise subtraction $b - s = c$, and the first nonzero entry in $c$ (from left-to-right) is positive.
Equivalently, following \cite{Munkres}:
\begin{align*}
    (a[{0}], a[{1}], \cdots, a[{n}] ) &\prec (b[{0}], b[{1}], \cdots, b[{n}]) \\
    \text{starting left-to-right,   if } a[i] < b[i] \text{ , else } a_i = b_i \quad \text{\&}  
    \quad a[i+1] &< b[i+1] \,\,(\text{for } 0 < i \le n)\quad.
\end{align*}

\subsubsection{tuple-tuple equality}

Suppose we have two tuples (of the same size, filled in with $n+1$ entries) $a, b \in \N^{n+1}$, then $a = b$ iff $a[i] = b[i]$ for all $i$ (the $i$th element of $a$ matches the $i$th element of $b$).

\subsubsection{tuples-to-numbers}
Next we provide a comparison is useful for mapping tuples to a number-system, while maintaining order (given two tuples, if one is greater than the other, the corresponding numbers will also obey the relationship).
Suppose for a number-system we have radix/base $r$. Then we may map a tuple to number in this system (1-1) by the inner-product (superscripts here are the exponential operation, i.e. power):
\begin{align*}
\text{num}_r(a) &= a[n] * r^{n} + a[n-1] * r^{n-1} + \cdots + a[1] * r^1 + a[0] * r^0\quad\quad .
\end{align*}
E.g. the binary number-system $(r=2)$, decimal ($r=10$), duodecimal system $(r=12)$, hexadecimal $(r=16)$. It can be shown if $a<b$, then their numbers base $r$ also comply with $\text{num}_r(a) < \text{num}_r(b)$.

\subsection{definition of lexicographic/dictionary order}




A list-of-tuples (the index-array) is said to be \textit{lexicographically}-ordered if all pairs-of-tuples comply with the tuple-tuple comparison $\prec$. Note this does not include equality, and hence all entries must be unique, we denote this order as \textit{sssort} or \text{well-ordered} in a mathematical sense. 
If we relax to equality in the comparison, we allow for duplicates and use the $\preceq$ comparison instead, this is denoted as \textit{ssort} order. 

\begin{figure}[h!]
\begin{tikzpicture}
    \draw (0,0) -- (12,0) -- (12,0.5) -- (0,0);
    
    \draw ( 0,-0.5) -- ( 2,-0.5) -- ( 2, 0.0) -- ( 0,-0.5);
    \draw ( 2,-0.5) -- ( 4,-0.5) -- ( 4, 0.0) -- ( 2,-0.5);
    \draw ( 4,-0.5) -- ( 6,-0.5) -- ( 6, 0.0) -- ( 4,-0.5);
    \draw ( 6,-0.5) -- ( 8,-0.5) -- ( 8, 0.0) -- ( 6,-0.5);
    \draw ( 8,-0.5) -- (10,-0.5) -- (10, 0.0) -- ( 8,-0.5);
    \draw (10,-0.5) -- (12,-0.5) -- (12, 0.0) -- (10,-0.5);
    
    \draw (0.0,-1.0) -- (0.5,-1.0) -- (0.5,-0.5) -- ( 0.0,-1.0);
    \draw (0.5,-1.0) -- (1.0,-1.0) -- (1.0,-0.5) -- ( 0.5,-1.0);
    \draw (1.0,-1.0) -- (1.5,-1.0) -- (1.5,-0.5) -- ( 1.0,-1.0);
    \draw (1.5,-1.0) -- (2.0,-1.0) -- (2.0,-0.5) -- ( 1.5,-1.0);
    \draw (2.0,-1.0) -- (2.5,-1.0) -- (2.5,-0.5) -- ( 2.0,-1.0);
    \draw (2.5,-1.0) -- (3.0,-1.0) -- (3.0,-0.5) -- ( 2.5,-1.0);
    \draw (3.0,-1.0) -- (3.5,-1.0) -- (3.5,-0.5) -- ( 3.0,-1.0);
    \draw (3.5,-1.0) -- (4.0,-1.0) -- (4.0,-0.5) -- ( 3.5,-1.0);
    \draw (4.0,-1.0) -- (4.5,-1.0) -- (4.5,-0.5) -- ( 4.0,-1.0);
    \draw (4.5,-1.0) -- (5.0,-1.0) -- (5.0,-0.5) -- ( 4.5,-1.0);
    \draw (5.0,-1.0) -- (5.5,-1.0) -- (5.5,-0.5) -- ( 5.0,-1.0);
    \draw (5.5,-1.0) -- (6.0,-1.0) -- (6.0,-0.5) -- ( 5.5,-1.0);
    \draw (6.0,-1.0) -- (6.5,-1.0) -- (6.5,-0.5) -- ( 6.0,-1.0);
    \draw (6.5,-1.0) -- (7.0,-1.0) -- (7.0,-0.5) -- ( 6.5,-1.0);
    \draw (7.0,-1.0) -- (7.5,-1.0) -- (7.5,-0.5) -- ( 7.0,-1.0);
    \draw (7.5,-1.0) -- (8.0,-1.0) -- (8.0,-0.5) -- ( 7.5,-1.0);
    \draw (8.0,-1.0) -- (8.5,-1.0) -- (8.5,-0.5) -- ( 8.0,-1.0);
    \draw (8.5,-1.0) -- (9.0,-1.0) -- (9.0,-0.5) -- ( 8.5,-1.0);
    \draw (9.0,-1.0) -- (9.5,-1.0) -- (9.5,-0.5) -- ( 9.0,-1.0);
    \draw (9.5,-1.0) -- (10.0,-1.0) -- (10.0,-0.5) -- ( 9.5,-1.0);
    \draw (10.0,-1.0) -- (10.5,-1.0) -- (10.5,-0.5) -- ( 10.0,-1.0);
    \draw (10.5,-1.0) -- (11.0,-1.0) -- (11.0,-0.5) -- ( 10.5,-1.0);
    \draw (11.0,-1.0) -- (11.5,-1.0) -- (11.5,-0.5) -- ( 11.0,-1.0);
    \draw (11.5,-1.0) -- (12.0,-1.0) -- (12.0,-0.5) -- ( 11.5,-1.0);
    
    \draw ( 14,-0.5) -- ( 16,-0.5) -- ( 16, 0.0) -- ( 14,-0.5);
    \node(draw) at (14, -0.75)   (a) {low \#};
    \node(draw) at (16, -0.75)   (a) {high \#};
    \node(draw) at (-0.5, 0.25)   (a) {$i$};
    \node(draw) at (-0.5, -0.25)   (a) {$j$};
    \node(draw) at (-0.5, -0.75)   (a) {$k$};
    
\end{tikzpicture}
\caption{\label{ssortdiagram} 
A cartoon depicting the ordering of a list-of-tuples array for a 3-index $A[i,j,k | I]$, along $i$ numerical values are sorted in ascending-order. Afterward $j$ is sorted in ascending-order within duplicate-values in $i$ (the previous column). Lastly, $k$ is sorted in ascending-order within duplicate-values in $i$ and $j$ (both previous arrays.}
\end{figure}
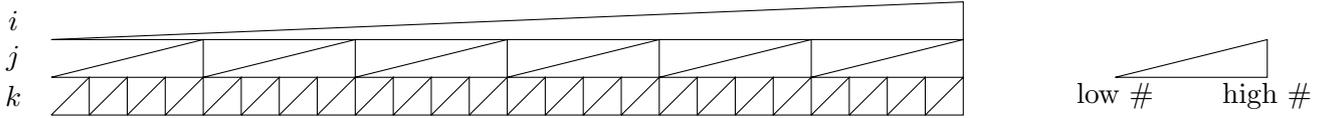









\subsubsection{lexicographic-sort algorithm}

Suppose we have a generic unsorted, potentially non-unique entry, list-of-tuples array (with column labels $\textbf{ 0, 1, 2, 3}, \cdots, \textbf{N}$):
\begin{align*}
    A\left[ \textbf{ 0, 1, 2, 3, }\cdots, \textbf{N} \,\,|\,I \,\right]\quad\quad.
\end{align*}
Next we introduce an auxiliary-array called the domain, $u$, which constrains the ranges of successive (column/index) sorts. For instance the sort is executed within successive elements in the domains array (range of list indices are [$u[n],u[n+1]$)), a constrained-sort:
\begin{align*}
    A\left[ \textbf{ 0, 1, 2, 3, }\cdots, \textbf{N} \,\,|\,\,u[n]\,:\,u[n+1]\,\,\right]\quad\quad.
\end{align*}

Upon sorting each index ($\textbf{0, 1, 2, 3}, \cdots, \textbf{N}$), the constrains on the sort are changed, and must increase. Initially there are no constraints. Then differences between the elements in the sorted index, form domains which constrains the sort of successive sorts.
Importantly, the domains $u^n$ must unionize with earlier domains $u^m$ ($m<n$) to respect earlier boundaries:
\begin{align*}
    u &= \bigcup_{\eta} \,u^{(\eta)}\quad\quad.
\end{align*} 
It would be convenient to have an \texttt{argsort} over small-domains, that can be parallelized over these small-domains.
This is quite natural for \texttt{mergesort} which partitions the data-set, before merging. Except we would like to specify this partition, which is potentially of different sizes. Then implement a sorting algorithm depending on the size of the partition: either \textit{Insertion-sort} xor \textit{Merge-sort}.

\subsubsection{computational time-complexity}


Let $N = M^n$ ($M$ is the length of a side of the dense-tensor), then we have to do successively constrained sort given by:
\begin{align*}
    \mathcal{O} &\sim
    M^n\left( \log{\left( \frac{M^n}{M^0} \right)} + \log{\left( \frac{M^n}{M^1} \right)} + \cdots + \log{\left( \frac{M^n}{M^n} \right)}\right) = \frac{n+1}{2} \,M^n\log{(M^n)} \\ &\sim \frac{n+1}{2} \,N\log{N}\quad\quad.
\end{align*}
Compare this to: $\mathcal{O}\sim n N\log N$, the independent sort of each column.

\subsubsection{theorems}


The following may be easily shown:
\begin{itemize}
\item Adding an arbitrary (random) column to the right sssort-order order respects the original sssort-order.
\item Removing an arbitrary-row (a tuple) maintains sssort-order.
\item Removing an arbitrary-column in sssort-order breaks it to ssort-order (creating the possibility of duplicates).
\end{itemize}

\acknowledgments

The author acknowledges the Arizona-State-University Department of Physics for support.





\bibliographystyle{fancybib}
\bibliography{refs.bib}

\end{document}